\documentclass[usenatbib, usegraphicx]{mn2e}

\usepackage{amssymb}

\title[Dynamical models of the elliptical galaxy NGC 4494. ]
{Dynamical models of the elliptical galaxy NGC 4494} 

\author[S.A.~Rodionov, E.~Athanassoula]
{S.A.~Rodionov$^{1,2,3}$\thanks{E-mail: seger@mail.ru}, 
E.~Athanassoula$^{1}$\\
$^{1}$ 
Laboratoire d'Astrophysique de Marseille (LAM), UMR6110,
CNRS/Universit\'e de Provence, Technop\^ole de Marseille-Etoile, \\
38 rue Fr\'ed\'eric Joliot Curie, 13388 Marseille C\'edex 20, France\\
$^{2}$Sobolev Astronomical Institute, 
St. Petersburg State University, 
Universitetskij pr.~28, 198504 St. Petersburg, Stary Peterhof, Russia\\
$^{3}$ Center de recherche INRIA Bordeaux -- Sub-Ouest, Bordeaux, 351
Course de la Lib\'eration, France \\ 
} 

\date{Accepted ???? ??? ??. Received ???? ??? ??; in original form ???? ??? ??}

\pagerange{\pageref{firstpage}--\pageref{lastpage}} \pubyear{20??}

\begin{document}
\label{firstpage}

\maketitle

\begin{abstract}
We present dynamical models of NGC 4494, which we built using our
iterative method presented in a previous paper. These models are
live $N$-body models consisting of equal mass particles, and they are steady
state as confirmed by a fully self-consistent evolution. Our goals were
twofold. The first one -- namely to test whether our iterative method
could indeed be used to construct galactic models following given
observational constraints, both photometric and kinematic -- was 
fully achieved. Our method allowed us to go beyond a simple spherical model
and to make full sets of rotating, axisymmetric models without any 
limitations to the velocity distribution. Our second goal was to understand
better the structure of NGC 4494, and more specifically to set constraints
on its halo mass. For this we tried three families of models: without halo,
with a light halo and with a heavy halo, respectively. 
Our models reproduce well the photometry and the kinematics, the latter
except specific regions where some non-equilibrium or non-axisymmetric
structure could be present in the galaxy (e.g. the kinematically decoupled
core). However, the lower order moments of the velocity distribution (up to
and including the second order) do not allow us to discriminate between the
three halos. On the other hand, when we extend the comparison to the higher
order moments of the velocity distribution obtained from the long-slit data,
we find that our light halo model fits the data better than the no halo, or
the heavy halo models. They also reproduce the shape of the angular
dependence of the PNe velocity dispersion in the outermost parts of 
the galaxy, but not the amplitude of its azimuthal variation. This may imply
that a yet more general class of models, such as triaxial, may be necessary
for a yet better fit.    

\end{abstract}

\begin{keywords} 
methods: numerical -- methods: N-body simulations -- 
galaxies: elliptical and lenticular -- galaxies: individual: NGC 4494 -- 
galaxies: kinematics and dynamics
\end{keywords}

\section{Introduction}
Dark matter around ordinary elliptical galaxies is one of the 
hottest topics in dark matter studies today. The main goal is to
obtain sufficient
constraints on the dark matter mass from observed stellar kinematics.
Traditional long-slit absorption line spectroscopy can only very
rarely give kinematics outside $2 R_e$, where $R_e$ is the effective
radius encompassing half the total light of the galaxy 
\citep[see for example, ][]{C09}. It is, nevertheless, possible to obtain 
line-of-sight velocities at larger radii using 
planetary nebulae (PNe), because  
their strong emission line at 5007 $\rm \AA$ [OIII] stands out against the faint
galaxy background. It is usually assumed that PNe trace the kinematics of
the underlying field stars\footnote{\citet{D05}, however, noted that
observations of PNe can be biased towards the kinematics of younger stars.}.
It is thus possible to obtain from the PNe the stellar
kinematic parameters at the periphery of the galaxy, out to $5-7 \; R_e$ 
\citep{G94, R03, D07, deL08, deL09, N09, C09}. 

\citet{R03} studied the kinematics in the outer part of three ordinary
elliptical galaxies, namely NGC 821, 3379 and 4494 (out to 4 -- 6 $R_e$) and 
found that their velocity dispersion profiles decline nearly
Keplerian-like at radii outside $2 R_e$. They modeled the
observational data by means of spherical Jeans models and by means of
orbit based models 
(see \citet{R03} for details) and they showed that their data are consistent
only with models with little or no halo. This result,
in good agreement with what was already found for NGC 3379 by
\citet{G94} and for NGC 4697 by \citet{Mendez01}, is very
surprising. Indeed, present theoretical and observational
data argue that ellipticals are formed from mergings of spirals, which
are known to have a considerable amount of dark matter
\citep[][and references therein]{Bosma04}. So, if the progenitors have
a considerable amount of dark matter, how can the merger product not have it?
Furthermore, this result is in conflict with the predictions of the standard
$ \rm \Lambda CDM$ cosmology.  

\citet{D05} constructed elliptical galaxy models from numerical
simulations of mergers of a pair of disk galaxies. 
Their resulting models have a ``normal'' massive dark halo and a 
velocity distribution with a high radial anisotropy in the outer
parts. The latter leads to a low observed (from most viewing angles)
velocity dispersion in the outer parts, which in turn leads to a
low estimated halo mass, contrary to the real dark halo mass of the model,
which is normal. In this way, \citet{D05} explain the results of 
\citet{Mendez01} and of \citet{R03} as due to the velocity
distribution in the merger remnant. 

Since this velocity anisotropy is
so crucial to the data interpretation, \cite{A05} examined whether
it was general, or whether it depended on the specific mergers
used. She examined 
the velocity anisotropy in multiple mergers, as would occur e.g. in
groups. In such cases, a pair merger is not examined as an isolated
event, but a whole sequence of mergers is considered. This model would
be more realistic in groups, but is also in good agreement with the 
standard $ \rm \Lambda CDM$ cosmology. The result
of such mergers are compatible with the observed properties of
elliptical galaxies \citep{W94, W96, A99}. Concerning the
velocity anisotropy in the outer parts, she found
that the result is more complex than the single pair mergers would
predict and that the anisotropy depends strongly on the time between two
successive mergers. Thus, more work is necessary to establish how
general the result of \citet{D05} is. A similar conclusion was reached
by  \citet{D07} 
who modeled the data of NGC 3379 and argued that there are
considerable discrepancies between the observations and
dark-matter-dominated simulations and re-iterate the question of
whether NGC 3379 has the kind of dark halo that the current
$\Lambda$CDM paradigm requires. 

\citet{deL08,deL09} constructed dynamical models of NGC 4697 and 3379 using
the $\chi^2$-made-to-measure particle method \citep{Syer96, deL07}
implemented in the NMAGIC code. Their main result is that the observational 
data are consistent with a fairly wide range of halo
mass profiles, although it was possible to place some limits on the halo mass. 
For NGC 4697, \citet{deL08} found that models with a low density halo 
with $v_c(5 R_e) \lesssim 200 \; km s^{-1}$  are not consistent with the
data, where $v_c(5 R_e)$ is the total circular velocity at $5 R_e$.
This, however, is a rather weak limit because even their model D with 
$v_c(5 R_e) \approx 210$ (see figure 15 in \citet{deL08}) has a very light
halo which contributes only about $35\%$ of the mass within $5 R_e$. 
For the galaxy NGC 3379
\citet{deL09} found that a model without a halo, as well as a model
with a heavy halo with $v_c(7 R_e) \gtrsim 250 \; km s^{-1}$ would be 
excluded by the observational data, but only at a $1\sigma$ confident level. 
So these constraints are not very strong.

The main problem in determining the halo mass profile from the
observed kinematics in elliptical galaxies is the
well known mass-anisotropy degeneracy.
The low velocity dispersion on the 
periphery of some elliptical galaxies can be
explained either by nearly isotropic models with a light halo or by radially
anisotropic models with a heavy halo. It is generally accepted that the
mass-anisotropy degeneracy can be broken by means of high order moments of
the line-of-sight velocity distribution \citep[LOSVD,][]{G93, vdM93}. As
shown in these papers, 
isotropic models have a Gaussian LOSVD and radially anisotropic models have
centrally peaked LOSVD as well as long tails. So these models can be
distinguished by means of high-order moments of LOSVD. However, 
the highly radial anisotropic models
presented in \citet{D05} have a LOSVD with relatively weak deviations from 
Gaussian \citep[see supplementary information in][]{D05}. 
Unfortunately, this means that, at least in some cases, breaking the
mass-anisotropy degeneracy  can be very difficult, if not practically
impossible. 

Let us also point out an obvious, but sometimes ignored, problem.
From a mathematical point of view, it is possible to prove the
existence of a given type of model by simply constructing dynamical models, 
but it is not possible to prove
its non-existence. Let us, for example, construct dynamical models of a real
galaxy by means of the
NMAGIC method or by means of our iterative method (see below).
More specifically, let us construct an axisymmetric model with some dark halo.
If this satisfies all observational data, then we have proven the
existence of a model with such a halo agreeing 
with the observational data. But if we fail to construct such a model, 
then formally we have not proven anything, since we can not exclude
that our failure is due to the method itself, or to the fact that
we have not searched sufficiently. We would have needed to prove that, 
if an equilibrium model with given parameters did 
exist, then our method would construct it.
This is not straightforwardly proven either for
our iterative method, or for the NMAGIC method, or for any other orbit based
method. Moreover, even if we did prove it, we would have only proven that an
axisymmetric model with this particular halo is excluded by the
observational data. We would not have proven anything about triaxial models, or 
about models with a somewhat different halo mass profile. 
Consequently, the conclusion of \citet{deL08,deL09} that the
observational data of NGC 4697 and 3379 are consistent with a wide
range of halo masses can be firmly believed. But, on the contrary, if
one finds that a set of models with a heavy dark halo do not agree
with the observational data, then one has at the best only an argument 
that the galaxies in question have a light dark halo, or no
halo. Unfortunately, this is no proof, since another, more general,
type of heavy halo could perhaps have fitted the observations. It is
thus very useful to try different approaches, to see 
whether this disagreement persists, or not. If more than
one method lead to the same conclusion, then the argument is
considerably strengthened.

Initially the present work was inspired by an article of
\citet[hereafter N09]{N09}, where the authors presented a large amount
of new observational data of the ordinary elliptical galaxy NGC 4494, 
resulting in
positions and velocities of 255 planetary nebulae out to seven effective
radii. They also presented new wide-field surface photometry from MMT/Megacam,
and long-slit stellar kinematics from VLT/FORS2. Using these data they
put constraints on the distribution of dark matter in this galaxy. They
constructed 
spherical dynamical models of the system using two different methods, but 
both of them are based on Jeans equations. They argue that some dark
matter is required by the data and their best fit model has a
relatively low halo mass. 
 
In \citet[hereafter RAS09]{RAS09} we presented an iterative method for
constructing equilibrium $N$-body models with given properties. This method
has been already widely used to construct initial conditions for
$N$-body simulations \citep[e.g.][]{RO08, M10} and can
also be directly used for constructing  
dynamical models of real galaxies from observational data. 
In contrast with NMAGIC models, our models consist of particles with equal
masses. They are steady-state and can be directly used in 
$N$-body simulations. 
For example, we can directly check that the constructed model is indeed in
equilibrium. Also, we can easily calculate for this model
any parameter which can be directly obtained from the positions and
velocities of the particles. 
 
In this article we will apply our iterative method to the
construction of
dynamical models of NGC 4494 from the observational data presented in N09.
We have two purposes. First, our intention is to demonstrate that our iterative
method can indeed be successfully used to
construct dynamical models of
real galaxies. The second aim is to return to the interesting question
of the halo mass of NGC 4494, using a different method from that of
N09, to see whether results can in any way depend on 
the method used. This is particularly important since the discrepancy
between the no-halo model of N09 and the observational data are not
large (see upper panels of their figure 12). 

We present the observational data in Sect.~\ref{s_observ} and our
method in Sect.~\ref{s_method}. In Sect.~\ref{s_models} we describe
our models and compare them to observations. We summarise and conclude
in Sect.~\ref{s_conc}.

\section{Preparation of observational data}
\label{s_observ}

We use the same observational data as N09. These include the surface
photometry, the stellar kinematics along the major and minor axes obtained by
means of long-slit spectroscopy, and velocities and positions of 255 planetary
nebulae (see N09). When constructing
our models, we use a physical system of units, i.e kpc and
$M_{\sun}$. We adopt a distance to NGC 4494 of 15.8 Mpc (see
N09), so that $1''$ is equal to $0.0766\, \rm kpc$. 
Our models can be easily rescaled to any other distance. 
Let us assume we have an equilibrium model constructed for some
distance $d_1$, and we want to rescale it to a distance $d_2 = C
d_1$. To keep the surface 
photometry unchanged we need to rescale all space coordinates of particles as
$r_2 = C r_1$. To keep the model in equilibrium we need to change the
mass of the model as $M_2 = C M_1$. Particle velocities need not be
changed, so all the observed kinematic parameters are unchanged. 
The new total luminosity of the galaxy
is $L_2 = C^2 L_1$, so that the new mass-to-light ratio is 
$\displaystyle \frac{M_2}{L_2} = \frac{1}{C} \frac{M_1}{L_1}$. 

Let us now describe how we prepare the observational data for use in our
iterative method. 

\subsection{Surface photometry}
\label{s_photometry}

We use the combined V-band surface photometry of NGC 4494 presented in
table A1 of N09.
These data are a combination of HST-based observations of \citet{L05},
ground based CCD observations of \citet{G94}, and the new observations
of N09. 

We, furthermore, make the following simplifications:
We assume that the ellipticity $\epsilon$ is the same at all radii and equal 
to 0.162, the mean value found in N09. 
We also assume that the shape of the isophotes is precisely elliptical,
so as not to introduce in the analysis unconstrained high order
isophote shape parameters. The first two rows of
table A1 in N09 give the surface brightness as a function of the  
intermediate axis $R_m$, measured in arcseconds.
This is related to the ellipticity and to the major axis $R_a$ 
by  $R_m=R_a \sqrt{1-\epsilon}$. We convert surface brightness in units
of $L_{\odot,V}/pc^2$ (assuming an absolute magnitude of the Sun in
the V-band $M_{\odot,V}=4.8$) and $R_m$
in parsecs using the adopted distance of $15.8$ Mpc. 

Excluding the innermost region  ($R_m < 5''$), this  surface
brightness profile is fitted very well by the S\'ersic law \citep{S68}
\begin{equation}
\label{eq_sersic}
I(R_m) = I_0 \exp{(-(R_m/a_s)^{1/n})}
\end{equation}
with parameters, $I_0=41764 \; L_{\odot,V}/pc^2$, $a_s = 0.008809 \;
\rm kpc$ and $n=3.3$, as shown by N09. 

Numerically, the surface brightness profile is described as follows: 
Inside the region $R_m < 0.46 \;\rm kpc$ ($\approx 6''$)  we 
interpolate the tabular data linearly. 
In the region $0.46 \; {\rm kpc} < R_m < 40 \;\rm kpc$ we adopt 
a S\'ersic profile.
In the region $40 \; {\rm kpc} < R_m < 50 \; \rm kpc$ we truncate 
the S\'ersic profile by means
of a 5-th order polynomial \citep[eq. 4]{D00b}. From the profile of the
surface brightness and the adopted value of ellipticity we can calculate
the two-dimensional distribution of surface brightness. The total 
luminosity of this model is
$L_V = 2.36 \times 10^{10} L_{\odot,V}$, or $M_V=-21.13$.
We assume that the stellar mass-to-light
ratio (M/L) is constant, in which case the distribution of surface
brightness equals the surface mass distribution to within an
unknown multiplier M/L.

\subsection{Kinematical data}
\label{s_kdata}

\subsubsection{Symmetries and system of coordinates}
\label{s_coordinates}

In order to prepare the kinematical data so that they can be used by
our iterative method, we need first to define a system of coordinates
which will be used for the model and to assume what symmetries the
galaxy has. 

In all the following we will assume that the galaxy is axisymmetric.
Elliptical galaxies can well be triaxial \citep{BT08}. Triaxial
models, however, have extra free parameters that add complexity to the
modeling and are beyond the scope of this paper.

Let us consider a Cartesian ($X$,$Y$,$Z$) coordinate system such that
the sky plane coincides with the $XZ$ plane. We choose the $Z$ axis so
that it coincides with the minor axis of the projected image of the
galaxy and then the $X$ axis will coincide with the major
axis\footnote {These are only 
the major and minor axes of the projected image, and 
the real principle axes of the galaxy can, of course, be different.}. 
We assume that the rotation axis is perpendicular to the $X$ axis
and therefore in the $YZ$ plane.
This
rotation axis is defined by its angle $\alpha$ with the $Z$ axis. 
If $\alpha=0$ then the rotation axis coincides with the
$Z$ axis, so the galaxy is edge-on.

We also assume that the galaxy
has a reflection symmetry with respect to the plane of symmetry
perpendicular to the rotation axis and centered on the center of
coordinates $(0,0,0)$. In that case, the 
observed image of the galaxy will have a reflection symmetry about
both the $X$ and the $Z$ axes. The observed line-of-sight velocity 
distribution will have a reflection symmetry about the $X$ axis. 
Also the observed line-of-sight velocity
distribution will have a reflection symmetry about 
the $Z$ axis, except for the velocity sign. 
It means that the points $(x,z)$ and $(-x,z)$ will have
the same line-of-sight velocity dispersion and that their 
line-of-sight mean velocities will be equal but with opposite sign. 
The velocity distribution in the
points $(x,z)$ and $(x,-z)$ will be fully identical. 
Consequently, if we know the
line-of-sight velocity distribution in any of the four quadrants, we know it
for the whole system. Similarly, if we assume such symmetry then 
all the observed kinematical data can be ``reduced'' to the first quadrant
of the sky plane ($x>0$, $z>0$). So in a first stage, we reduce all 
kinematical data to the first quadrant. 
 
In our iterative method the input kinematical data should be given as
mean velocities and velocity dispersions in a set 
of two-dimensional areas on the sky plane and we need to present the
observed velocities in this manner.
  
\subsubsection{Long-slit spectra kinematics}
\label{s_longslit}

Part of our kinematical data are obtained by means of long-slit spectroscopy
\citep{C09} from spectra taken along the
major and the minor axes of the galaxy.
The width of the slit was $1''$. Thus we have along each axis profiles of
the rotation, of the velocity dispersion, as well as two Gauss-Hermite
moments ($h_3$ and 
$h_4$, see appendix~\ref{s_gh}). When constructing the models we
will use the rotational velocity profiles along the major axis and of the
velocity dispersion along both principal axes (while the Gauss-Hermite
moments will be used for analysis of the so constructed models, see
Sect.~\ref{s_models}).
We use these data almost ``as is'' without any parametrization. We only 
bin the data suitably.

 Let us describe how we prepare the profile of the mean velocity along 
the major axis. 
From \citet{C09} we get the table containing values of the
mean velocity for different points along the
major axis. So we have a set of pairs $x_i$, $v_i$,
where $x_i$ is the coordinate of the
data point on the major axis and $v_i$ is the observed
line-of-sight mean velocity at this point. At first we ``reduce'' these data
to the first quadrant (see previous Sect.). For the mean velocity
this implies multiplying $x_i$ and $v_i$ by -1 for each data  
point with $x_i<0$. We need to convert these data 
into a set of two dimensional areas 
with known mean velocity. These long-slit data were obtained
from rather narrow zones along the major axis. We assign these data to
a wider zone. We assign the data along the major axis to the area defined by 
$x < 0.5'' \, || \, \varphi < 10^{\circ}$, where $\varphi$ is the angle
between the
current radius vector and the $x$-axis measured on the $XZ$
plane, and $||$ is a logical OR. We divide this area along the
$x$-axis into the pieces so that each long-slit data
point corresponds to a single piece. We do this as follows. 
We sort the data points by 
$x_i$. For each data point we define two values as 
$l_i = (x_{i - 1} + x_{i})/2$ and $u_i = (x_{i} + x_{i+1})/2$. We set
$l_1 = 0$ and $u_n = 2 x_n - u_{n-1}$, where $n$ is the number of
data points. For each data point we assign a two-dimensional area defined as 
$x>l_i \, \&\& \, x < u_i \, \&\& \, (x < 0.5'' \, || \, \varphi <
10^{\circ})$, where $\&\&$ is a logical AND.
We then join some of these areas in the following way. We create an
$N$-body system with a surface distribution of particles obtained from 
the surface
photometry (see Sect.~\ref{s_photometry}). The number of particles is 
$N=500\,000$. For each area we calculate the number of particles which are
situated in this area. 
If the number of particles in some area is less than $1000$
then this area is joined with a neighbouring area. 
The value of the velocity in each composite area is calculated as 
the mean value of the velocities of the
constituent areas weighed by the number of particles in each area.  
This way of binning the observational
data is rather unusual but it is convenient for us, because we make sure that
in each area there is a sufficient number of particles.

The other profiles are prepared in the same way. The data along the minor axis
are assigned to the area defined as $z < 0.5'' \, || \, \varphi > 80^{\circ}$. 
  
 In our algorithm we don't use any information on the errors of the
 data. But we will use binned long-slit data for figures, so, in order
 to plot error bars, we need to calculate the errors of the binned data. 
We calculate the error of a binned datum as 
$e = \frac{1}{n^{3/2}}\sum e_i$,
where $n$ is the number of original data points from which binned datum
was calculated, and $e_i$ are the errors of these data points.

\subsubsection{PNe kinematics}
\label{s_pn}

\begin{table*}
\centering
\caption{Parameters of the five planetary nebulae groups. The first column
gives the name of the corresponding area (see Fig.~\ref{fig_pn_areas})
and the second and third ones give its lower and the upper azimuthal
boundaries ($\varphi^{\rm (min)}$ and $\varphi^{\rm (max)}$,
respectively). The remaining columns give parameters relevant to the 
group of PNe in the corresponding area. Here $n$ is the number of PNe, 
$R_{xz}^{\rm (min)}$ and 
$R_{xz}^{\rm (max)}$ are the minimal and the maximal
distance to the galaxy center, $\bar V$ is the mean
velocity, $\Delta \bar V$ is the error of the mean velocity,
$\sigma$ is the velocity dispersion (standard deviation) and 
$\Delta \sigma$ is error of the velocity
dispersion. } 
\label{t_pn}
\begin{tabular}{c||ccccccccc}
\hline
Area & $\varphi^{\rm (min)}$ & $\varphi^{\rm (max)}$ & 
n & $R_{xz}^{\rm (min)}$  
& $R_{xz}^{\rm (max)}$ & $ \bar V$ & $ \Delta \bar V$ & $\sigma$ 
& $\Delta \sigma$ \\
& & & & [arcsec] & [arcsec] & [$\rm km/s$] & [$\rm km/s$] & [$\rm km/s$] 
& [$\rm km/s$]\\
\hline
 
A1 & $0^{\circ}$  & $18^{\circ}$ & 11 & 166.3 & 297.1 & 72.7 & 15 & 49.7 &
11.2 \\
A2 & $18^{\circ}$ & $36^{\circ}$ & 18 & 126.7 & 291.8 & 14.1 & 19.2 & 81.3 &
14 \\
A3 & $36^{\circ}$ & $54^{\circ}$ & 17 & 131.1 & 378.9 & 21.9 & 25.5 & 105 &
18.8 \\
A4 & $54^{\circ}$ & $72^{\circ}$ & 14 & 132 & 277.6 & -2.64 & 24.5 & 91.8 &
18.2 \\
A5 & $72^{\circ}$ & $90^{\circ}$ & 14 & 133.4 & 269.8 & -24.7 & 14.3 & 53.4
& 10.6 \\

\hline
\end{tabular}
\end{table*}

\begin{figure}
\begin{center}
\includegraphics[width=7.5cm,angle=-90]{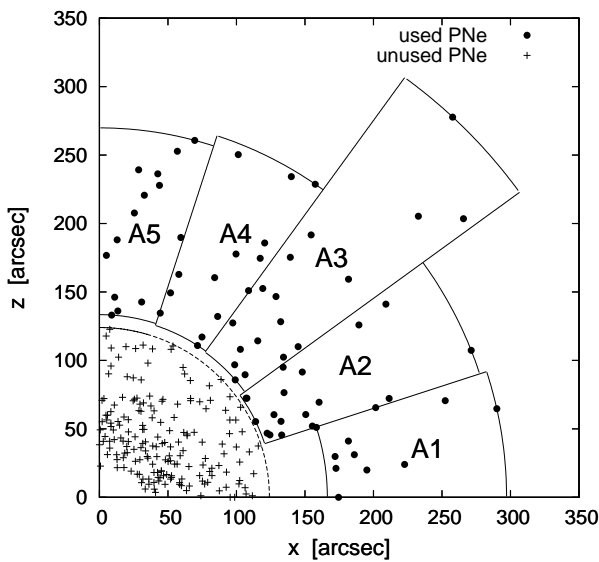}
\end{center}
\caption{Five PNe areas. All planetary nebulae have been moved into the first
quadrant.}
\label{fig_pn_areas}
\end{figure}

We use the PNe data in the outer part of the galaxy, which long-slit data 
can not reach. More precisely, we use all PNe whose distance from the 
galaxy center $R_{xz}=\sqrt{x^2+z^2}$ is larger than $124''$
($9.5$~kpc). 

We ``reduce'' the PNe data to the first quadrant 
(see Sect.~\ref{s_kdata}). 
We divide the first quadrant into five zones with an equal opening
angle $\varphi$ and use the $\varphi$ angle of the individual PNe to place them
into the appropriate zone. The
parameters of these five PNe groups are shown in
table~\ref{t_pn}. For each group we define an area 
($\varphi^{\rm (min)}<\varphi<\varphi^{\rm (max)}) \, \&\& \, 
(R_{xz}^{\rm (min)} <R_{xz}< R_{xz}^{\rm (max)})$. 
These areas are shown schematically in Fig.~\ref{fig_pn_areas} and we
will refer to these areas as A1, A2, A3, A4, A5. 

For each group (area) we calculate the mean velocity and the velocity
dispersion with the corresponding errors. We note that here we take
the standard deviation as the dispersion. For the calculation of the
standard deviation we use an
unbiased estimator \citep[p. 171]{K51}, although
for such relatively large samples ($n>10$, see
table~\ref{t_pn}), the use of an unbiased estimator is not essential.

We want to draw attention to two features of the PNe
velocity distribution. The first feature is that the 
velocity dispersion at intermediate angles is noticeably higher than the
velocity dispersion along the major or the minor axes. The velocity
dispersion in area A3 is twice as high as in areas A1 or A5. 
This feature can be clearly seen on the right panel of figure~7 in
N09, where, except for the inner areas the iso-dispersion contours are
elongated in a direction intermediate between the two principal axes.
Since this figure was constructed under the assumption of triaxial
symmetry, this peculiarity remains true independent of the
axisymmetric assumption. The second feature 
is that the value of the mean velocity in
A5 is negative and that it is low in the intermediate areas A2, A3, and A4. 
This feature can be explained by a twisting of 
the rotation axis in
the outer part of the galaxy (see left panel on figure~7 in N09 and
corresponding discussion), but can not be reproduced by the models with an
axisymmetric velocity distribution. We, therefore, can not take into
account in the iterative process the mean velocity from all
regions. Since in axisymmetric systems there is no rotation along the
minor axis and since the value from the A1 area agrees very well with
the long-slit data (figure 3.d), we will choose to use the mean velocity
from area A1.

\section{Method}
\label{s_method}

\subsection{General outline}
\label{s_methodgen}

We want to construct an equilibrium $N$-body model of an elliptical
galaxy from its observational data. As described in the previous
section, for the galaxy under consideration we
have surface photometry, a distance estimate and
various line-of-sight kinematics. Assuming that the 
stellar mass-to-light ratio (M/L) is constant, we can obtain from the
surface photometry and the distance the
surface mass distribution to within an unknown multiplicative constant M/L. 
In the case of an $N$-body model, this implies that we have the
projected surface distribution of particles, but the mass of the
individual particles is not known (in our models all particles 
have the same mass). We note, however, that the M/L can not be arbitrary, 
because it is related to the line-of-sight 
velocity dispersion in the central part of the galaxy which we know from 
observations (see Sect.~\ref{s_ML}). We also assume that the 
galaxy is axisymmetric. Observations do not give us the inclination of 
the rotation axis, so this is a free parameter. 

As result, our task is to construct an equilibrium $N$-body model with
the given projected surface distribution of particles and the given
line-of-sight kinematics.
The total mass of the model is unknown (M/L is unknown), but 
should be found somehow. And the model should be axisymmetric with a
given axis of rotation. The last condition is of course optional, but
simplifies the modeling.

\begin{figure}
\begin{center}
\includegraphics[width=8cm]{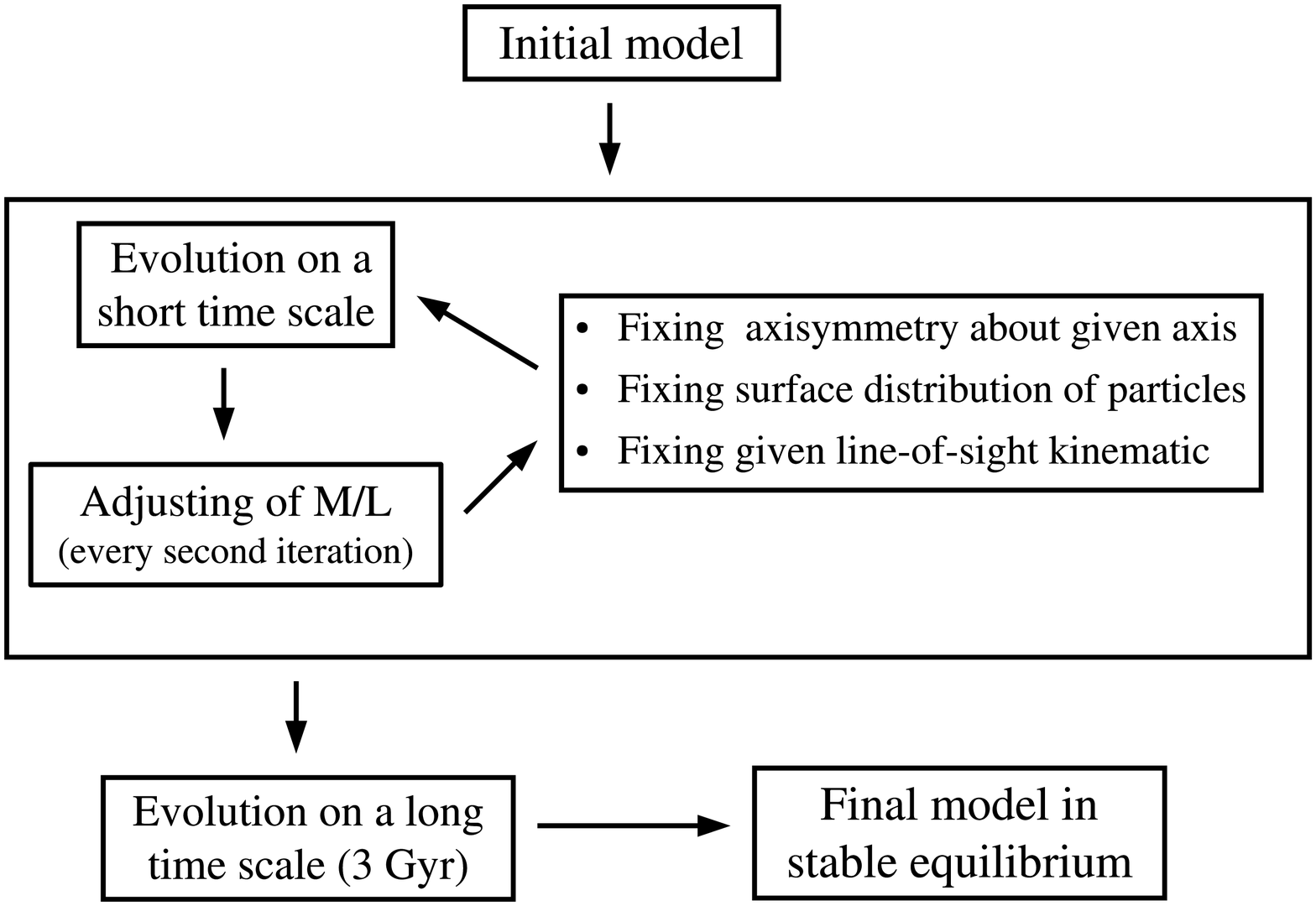}
\end{center}
\caption{The scheme of the iterative method which we use for constructing
the $N$-body model of an elliptical galaxy from observed data. 
}
\label{fig_scheme}
\end{figure}

In RAS09 we presented an iterative method for constructing
equilibrium $N$-body models with given properties. The idea
of the iterative method is very simple. It relies on constrained, 
or guided, evolution. We simply evolve the system while 
constraining the desired system properties (see RAS09).
The same conceptually method, with only relatively minor modifications can
be applied to our present task. 

The scheme of the modified iterative method
which we use in this work is outlined schematically in Fig.~\ref{fig_scheme}.
We will first briefly describe the whole method and then each
part of the algorithm in detail. We start by building an initial $N$-body
model which has a rigid halo and the desired projected surface distribution of 
the particles. The distribution of particles along the line-of-sight can be
arbitrary. The velocities of the particles and the total
mass of the system (M/L) can also be arbitrary. 
This model will be the starting point for the iterative procedure.
At the start of each iteration we calculate the evolution of the system over a
short time. Then we put the model through a specific
procedure which adjusts the total mass of the system (see
Sect.~\ref{s_ML}) every second
iteration. We then, if necessary, impose the condition of axisymmetry
about the given axis. We do this in the usual way, i.e. by randomizing the
particle azimuthal
angles. Next, we fix the surface distribution of the particles
(see Sect.~\ref{s_fixsd}). Finally, we fix the line-of-sight
kinematics to agree with the observations (see
Sect.~\ref{s_fixvlos}). We repeat this iteration cycle a number of
times until the velocity distribution and the total mass of the system  
stop changing. The model at this stage is already in equilibrium, or
very close to it. To be sure of this, we calculate the evolution of
the system over a long time scale (3 Gyr in this article) after which
we obtain the final model in stable equilibrium. Let us stress
that we consider this long time scale evolution only as a part of our
algorithm for constructing equilibrium models.

In the axisymmetric case we assume that the galaxy 
rotates about the axis which lies in the $Z-Y$ plane. This
rotation axis is defined by its angle $\alpha$ with the $Z$ axis. 
If $\alpha=0$ then the rotation axis coincides with the
$Z$ axis, so the galaxy is edge-on.

Having described the general outline of our method, we will now
describe each individual step.

\subsection{Fixing the projected surface distribution of particles}
\label{s_fixsd}

Here we will describe how we fix the projected surface distribution of
particles. We do this using a
method very similar to that described in RAS09 for fixing the particle
mass distribution. The idea of that algorithm is as follows: We start
with a $N$-body system,
which is the result of a short evolution (Fig.~\ref{fig_scheme}) and to
which we will refer to as the ``old'' system. We need to fix the mass
distribution in this model according to given constraints. 
We create a ``new'' $N$-body system with the desired mass distribution,
and we ``transfer'' the velocity distribution from the
``old'' to the ``new''
model. The basic idea of our velocity transfer algorithm is very simple.
We assign to the new-model particles the velocities of
those particles from the old model that are nearest to the ones in the new
model. This algorithm is described in detail in RAS09 (section 2.4). We note
that this algorithm has a free parameter $n_{nb}$ --- the number of neighbours.

It is very easy to modify this algorithm to fix only 
the projected surface distribution of particles. We need to construct 
a ``new'' $N$-body system with a
given surface distribution of particles. In the new model 
only the $x$ and $z$ coordinates of the particles are defined. The
$y$-coordinates,  the velocities and the
mass of the particles should be carried over from the old model. 
The total mass of the new system is set
equal to the total mass of the old system. We note again that in our
models all particles have the same mass.
In the algorithm described in RAS09
we ``transfer'' from the old to the new model only the velocities. 
In the present case we need to ``transfer'' also the y-coordinates of
the particles. When we search for the nearest particle we need to do so
in the two-dimensional space $X-Z$. The
$y$-coordinate should not be taken into account because we copy it from
the old model particle to the new model particle together
with the velocities.

\subsection{Fixing the line-of-sight kinematics}
\label{s_fixvlos}

Let us first describe an
algorithm for fixing the line-of-sight mean velocity
for the case where we do not assume any symmetry in the system.
Our task is as follows. We have an $N$-body system and some two
dimensional area on the sky plane where we need to fix the
line-of-sight mean velocity to the observed value. In our
models the sky plane coincides with the $XZ$ plane, so the 
line-of-sight velocity is the
velocity along the $Y$-axis. The given 
area is a two dimensional area in the $XZ$ plane, so when we search the
particles which belong to the given area we do it regardless 
of $y$-coordinates of the particles. We denote by $\bar v_y$ the
desired value of the line-of-sight mean velocity in the given area and
by $\bar v_y^{\prime}$ the mean value of the $y$ velocities of all
particles in the area. We need to change somewhat the particle
velocities so that $\bar v_y^{\prime}$ becomes equal to $\bar v_y$.
This is achieved by setting the new $y$ velocity
component of particle $i$ in the given area to
\begin{equation}
v_{y i} = v_{y i}^{\prime} + (\bar v_{y} - \bar v_{y}^{\prime}) \,,
\end{equation}
where $v_{y i}^\prime$ is the current value of the $i$-th particle
and $v_{y i}$ is the corrected $i$-th particle $y$ velocity.

 The algorithm for fixing the line-of-sight velocity dispersion is
 very similar. Let us denote by $\sigma_y$ the desired value of the
 line-of-sight velocity dispersion in the area under consideration. In
 the given area we calculate the 
current value of the line-of-sight velocity dispersion $\sigma_y^{\prime}$
and the current value 
of the line-of-sight mean velocity $\bar v_y^{\prime}$.
The new $y$ velocity
component of particle $i$ in the given area is set as
\begin{equation}
v_{y i} = (v_{y i}^{\prime} - \bar v_y^{\prime})
\frac{\sigma_y}{\sigma_y^{\prime}}
+ \bar v_y^{\prime} \, .
\end{equation}

Let us now consider a galaxy with the following symmetries (as described
in Sect.~\ref{s_kdata}). The galaxy is 
axisymmetric, with a symmetry axis which lies in the
$YZ$ plane, and also has a reflection symmetry with a
plane of symmetry perpendicular to the axis
of symmetry and containing the center of coordinates $(0,0,0)$. 
In this case, if we know the
line-of-sight velocity distribution in the single
quadrant then we know it for the whole system (see Sect.~\ref{s_kdata} for
details). 
So, if we assume such a symmetry then all observed
kinematical data can be ``reduced'' to the first quadrant ($x>0$, $z>0$). 

In the case of such a symmetry the task of fixing the line-of-sight kinematics
is as follows. We have some two
dimensional area in the first quadrant of the sky plane for which we have
the line-of-sight mean velocity and/or 
the line-of-sight velocity dispersion. We
need to fix these kinematic parameters in the given area taking into account
the discussed symmetries. We invert the sign of $v_y$ for each particle with
$x<0$ and then we flip all particles to the first quadrant, i.e. we set
$x=|x|$ and $z=|z|$ for each particles. Now we can apply the
algorithms for fixing the line-of-sight kinematic parameters which we
described in the first part of this section. We then flip all particles
back to their original positions and we invert the sign of $v_y$ for
each particle with $x<0$. 

\subsection{Adjusting the total stellar mass}
\label{s_ML}
Since we have no a priori knowledge of the total stellar mass in
the galaxy, we have to construct our models assuming that the stellar $M/L$ is
unknown. In our case this implies that the total mass of the system is
unknown. This of course raises questions. Does an equilibrium model
with properties in agreement with observations exist for a unique
value of the mass? Or for a range of values? For example, for a given 
equilibrium, spherical, isotropic model with known projected surface
distribution 
of particles and line-of-sight velocity dispersion in the center of the 
model, the total mass of the model would be uniquely
determined. But in our case it is not so obvious. Moreover, as we will
show, models with different inclinations of the rotation axis have slightly 
different total masses (see Table~\ref{t_ML}, for example models AL0 and
AL45). So if we don't fix the inclination of the rotation axis then the
total mass may well not be unique.

 Nevertheless we need to find some value of the total mass for which we can
construct the equilibrium model. The straightforward way to solve this
problem is to construct a series of models with different
total masses and choose the one which, according to some definition,
is closest to equilibrium, or to find ranges of values for which the
resulting equilibrium model is in agreement with the observational
constraints. This, however, would be excessively time consuming, due
to the large number of models that need to be constructed.  

We will, therefore, use a different algorithm, in which we adjust the total
mass during the iterative process (Fig.~\ref{fig_scheme}). 
As we described earlier, in each iterative step we let the
system evolve on a short
time scale (Fig.~\ref{fig_scheme}). We calculate in the beginning and in
the end of this evolution the velocity dispersion along the
line-of-sight in some given part of the system, which we denote by
$\sigma_1$ and $\sigma_2$, respectively.
In all experiments described here this given part was a
sphere with radius equal to $10$~kpc. 
We note that the value of $\sigma_1$ is
partly defined by the given line-of-sight velocity dispersion which we fix on 
the previous stage of the iteration (see Fig.~\ref{fig_scheme}). On the other
hand the value of $\sigma_2$ is influenced by the total mass of the model.
So, if we choose a value of total mass which is not
appropriate and we
don't change it during the iterative procedure,
then $\sigma_1$ and $\sigma_2$ will be different after
any number of iterations. We want to construct the equilibrium model so 
that the values of $\sigma_1$ and $\sigma_2$ are close, so
we adjust the total mass in the system by multiplying all
velocities by a factor of $\sigma_1/\sigma_2$ and the masses of
all the particles by a factor of $\sigma_1^2/\sigma_2^2$. By trial and
error we found that the iterations converge faster if 
we adjust the total mass not at every iteration but only every second
iteration.   

For a model without dark halo this means that, if the system at 
the end of the evolution was in equilibrium, then the rescaled 
system will also be in equilibrium, but will have the 
line-of-sight velocity dispersion in the selected
part of the model equal to $\sigma_1$. For models with halo this explanation
is of course not valid, but this is not a serious drawback since there
will be further iterations to bring the system to equilibrium. We thus
used this algorithm for constructing all
our models, including models with dark halo, and it
worked well in all cases. I.e. all these models 
constructed by means of the iterative method with this 
mass-adjusting algorithm were in equilibrium. 
But we note that it is possible that, for models with a very massive
halo dominating the central 
part of the galaxy, this algorithm may not work and the iterations
would not converge. In such cases one would have to resort to the
straightforward but very time consuming algorithm described above. 

\section{Models}
\label{s_models}

\subsection{Description of the models}

We constructed models with three types of halo. The first type is models
without halo. The second and third type are rigid NFW halos \citep{N96,N97}
with density profile

\begin{equation}
\rho(r) = \frac{\rho_{\rm s}}{(r/r_{\rm s})(1+r/r_{\rm s})^2} \, ,
\label{eq_NFW}
\end{equation}
where $\rho_s$ and $r_s$ are the characteristic density and scale radius of
the halo. The second type was found in N09 to be the best-fitting NFW
halo for this galaxy (see Sect.
4.2.5 in N09). This is a relatively light halo with parameters 
$\rho_s = 0.0019 \; M_{\odot} pc^{-2}$ and $r_s = 32 \; \rm kpc$, i.e.
a concentration parameter $c_{\rm vir} \approx 8$ and a virial mass
$M_{\rm vir} \approx 10^{12} M_{\odot}$ (see N09). The third 
type is a relatively massive halo with parameters 
$\rho_s =  0.00522 \;  M_{\odot} pc^{-2}$ and $r_s = 26.5 \;\rm kpc$. 
N09 found that a model with such massive halo is a relatively poor fit
of the observational data (see
their Sect. 4.2.5). Since we use a totally different approach, we include this
model to test whether this conclusion is method-dependent, or not. 
This halo has $c_{\rm vir} = 12.3$ and 
$M_{\rm vir} = 2 \times 10^{12} M_{\odot}$.  
 
Another parameter of our models is the
angle $\alpha$ defining the inclination of the
rotation axis (see Sect.~\ref{s_method}). We construct models for two
values of the angle $\alpha$, namely $0^{\circ}$ and $45^{\circ}$. 

As we discussed in Sect.~\ref{s_pn} the azimuthal variation of the
PNe velocity dispersion presents an interesting feature in the outer parts.
Namely, the velocity dispersions in areas
close to the principle axes (areas A1 and A5) are considerable smaller than
in intermediate areas (areas A2, A3 and A5). It is not clear whether
such a feature can be reproduced by an equilibrium
axisymmetric system. So we construct two sets of models.
In the first ones, which we denote as ``A'',
we don't try to model this feature. 
Thus, we ask the iteration method to fit the following
quantities:
\begin{itemize}
\item 
The projected surface distribution of the particles (see Sect.~\ref{s_photometry})
\item 
The radial profile of the mean velocity along the major axis obtained from
long-slit spectroscopy
\item 
The radial profiles of the velocity dispersion along the major and minor axes
obtained from long-slit spectroscopy (see Sect.~\ref{s_longslit})
\item 
The mean velocity in area A1 obtained from PNe kinematics. 
Note that we don't fix the mean velocity in other areas (see Sect.~\ref{s_pn})
\item 
The velocity dispersion in the two PNe areas A1 and A2, which
  are close to the major and minor axes, respectively.
\end{itemize}

For the second set of
models, which we denote by ``B'', we take into account all available
information. We, therefore, try to fit the velocity dispersion in all
five PNe areas separately.
This introduces a considerable extra difficulty because of the peculiar
PNe velocity distribution.
As already mentioned, such a fit may of course not be possible with models
such as ours, 
i.e. with models that are both axisymmetric and in equilibrium. It is,
nevertheless, important to try for at least two reasons. First it is
useful in all cases to try and fit all available data, since only that can
tell us how far off the attempted fit is from reality. Second, since
we allow the model total freedom regarding the velocity anisotropy, a
fit may be found, in which case an interesting effect of anisotropy
would be revealed.

We will, therefore, discuss in total 12 models, which we will denote as follows.
The first symbol in the name of the model denotes the set (A or B) and the
second denotes the halo. Here ``N'' is for models without halo, ``L''
is for models with a relatively light NFW halo
($\rho_s = 0.0019 \; M_{\odot} pc^{-2}$, $r_s = 32 \;\rm kpc$) and 
``H'' is for models with a relatively massive NFW halo 
($\rho_s =  0.00522 \; M_{\odot} pc^{-2}$, $r_s = 26.5 \;\rm kpc$). The
last number in the model name denotes the angle $\alpha$ of 
the rotation axis inclination.

Let us now discuss in some detail how we construct these
models. There is considerable freedom in choosing the initial model,
from which the iterative search will start. Our basic initial model is a model
with the given surface distribution of particles (generated using
the rejection method), and with zero velocities. The $y$
coordinates of the particles were chosen as random numbers 
from the interval $[-1,1]$ and the initial mass was chosen
according to $M/L_V = 1$. These choices are of course totally {\it ad
  hoc}, but this does not matter since other values, although starting
the iterative procedure (see Fig.~\ref{fig_scheme}) from a different
initial model, lead to essentially the same 
final model. 
In practice, there is only one significant property of
this initial model, namely whether it is rotating (or
not). Indeed, if we start the iterative search from a rotating, rather
than a non-rotating initial model, we will end up with a model that is
different, albeit not in all properties. 
A rotating initial model can be constructed as follows. We first take
the basic non-rotating initial model and put it in  
the iterative procedure (see Fig.~\ref{fig_scheme}).
After a relatively small number of iterations, the model becomes 
close to steady state. We then choose the axis of rotation and set,
for all particles, the azimuthal velocity  
with respect to the chosen axis equal to the circular velocity. We
have thus obtained a rotating initial model.

Models with $\alpha=45$ constructed from rotating and from
non-rotating 
initial models are practically identical. Non-inclined models ($\alpha=0$),
however, are slightly different. In particular, there is a difference in
rotation in areas far from major axis. We note that we fix the mean
line-of-sight velocity only in areas close to major axis (long-slit data
along major axis and mean velocity in A1 area).
It is not surprising that models constructed from initially rotating
models rotate slightly faster in areas far from major axis.
But this difference is not sufficiently significant to warrant further
discussion. 

 Here we will discuss models constructed from the rotating initial model. 
Each model was constructed as follows. 
At first we construct a basic non-rotating initial model with 
$N=300\,000$ particles. We put this model into the
iterative procedure and make $100$ iterations with relatively low
precision. This is possible, because each iteration is very short and
errors do not accumulate (RAS09). 
We use the fast $N$-body code gyrfalcON \citep{D00, D02} with an 
integration step and a softening length equal to
$dt=1 / 2^{12}$ Gyr and $\epsilon=0.05$ kpc, respectively\footnote{We use
a system of
units where the unit of length is $u_l=1 \;\rm kpc$, the unit of
velocity is $u_v=1 \;\rm km/sec$, the unit of mass is 
$u_m = 10^{10} \;\rm M_{\odot}$ and consequently 
the unit of time is $u_t \approx 0.98 \;\rm Gyr$. For simplicity, all
time values in this papers are presented with the 
assumption that $u_t=1 \;\rm Gyr$.}. The tolerance
parameter for gyrfalcON was set to $\theta_t=0.9$ and the duration of
each iteration to $t_i=0.05$ Gyr. Using this constructed model, we
create the rotating initial model with $N=500\,000$ particles. 
Again we put this model into the iterative procedure and make $500$
iterations. The integration step and the softening length were taken
$dt=1 / 2^{14}$ Gyr and $\epsilon=0.02$ kpc, respectively. The
duration of each iteration and the
tolerance parameter for gyrfalcON were chosen as in the previous
stage. The final stage of our procedure is a free evolution over a
time scale of  3~Gyr (see~Sect.\ref{fig_scheme}), with parameters
$dt=1 / 2^{15}$~Gyr, $\epsilon=0.02$ kpc and $\theta_t=0.6$. We would
also like to mention that before fixing the parameters to these
values, we made a number of tests, such as increasing the number of
particles up to tenfold, and did not find any significant improvements
in the fits.

\begin{table}
\centering
\caption{Stellar M/L and relative halo mass for the
constructed models. The first column gives the name of the model, the
second and third columns the stellar mass-to-light ratios in $V$ and
$B$ bands, $M/L_V$ and $M/L_B$, respectively. The fourth and fifth
columns give the ratio of the halo mass, $M_h(x)$, to the stellar mass,
$M_{\ast}(x)$, both calculated within a sphere of radius $x = R_e$ and
$x = 5 R_e$, respectively.
}
\label{t_ML}
\begin{tabular}{ccccc}
\hline
Model & $M/L_V$ & $M/L_B$ &  $ \displaystyle \frac{M_h(R_e)}{M_{\ast}(R_e)}$
      & $ \displaystyle \frac{M_h(5 R_e)}{M_{\ast}(5 R_e)}$ \\  
\hline
AN0   & 4.16 & 4.82 & 0 & 0     \\  
AL0   & 3.55 & 4.11 & 0.13 & 0.96  \\  
AH0   & 2.76 & 3.22 & 0.36 & 2.52  \\  
AN45  & 4.23 & 4.90  & 0    & 0     \\  
AL45  & 3.73 & 4.32 & 0.12 & 0.90  \\  
AH45  & 3.15 & 3.65 & 0.30 & 2.18  \\  
BN0   & 4.21 & 4.89 & 0  & 0     \\  
BL0   & 3.55 & 4.12 & 0.13 & 0.96  \\  
BH0   & 2.89 & 3.35 & 0.35 & 2.41  \\  
BN45  & 4.27 & 4.95 & 0    & 0     \\  
BL45  & 3.81 & 4.41 & 0.11 & 0.88  \\  
BH45  & 3.19 & 3.69 & 0.30 & 2.16  \\  
\hline
\end{tabular}
\end{table}

The stellar mass-to-light ratio and relative mass of 
the dark halo inside one and five effective radii, respectively,  
are shown in Table~\ref{t_ML}. According to N09 the value of
the effective radius $R_e = 48''.2 \approx 3.69 \, \rm kpc$.
We use photometry in the V-band, so for us it is
more straightforward to calculate the
mass-to-light fraction in V-band $M/L_V$. To compare our results with N09 we
also calculate mass-to-light fraction in B-band $M/L_B$. We assume that
$(B-V)$ is 0.65 for the Sun and 0.81 for NGC 4494 (see N09). We
note that, if we change the adopted distance, the mass-to-light will
be changed (see Sect.~\ref{s_observ}). For all models, there is
considerably less halo mass than stellar mass within $1 R_e$.
For models with a ``light'' halo, the mass of the dark matter inside $5 R_e$ 
is approximately equal to the mass of the stellar component, while
for model with a ``heavy'' halo the mass of dark matter
inside $5 R_e$ is more than twice that of the stellar component.

\subsection{Discussion of the first set of models} 

\subsubsection{Basic comparison} 

\begin{figure*}
\begin{center}
\includegraphics[width=17.3cm,angle=-90]{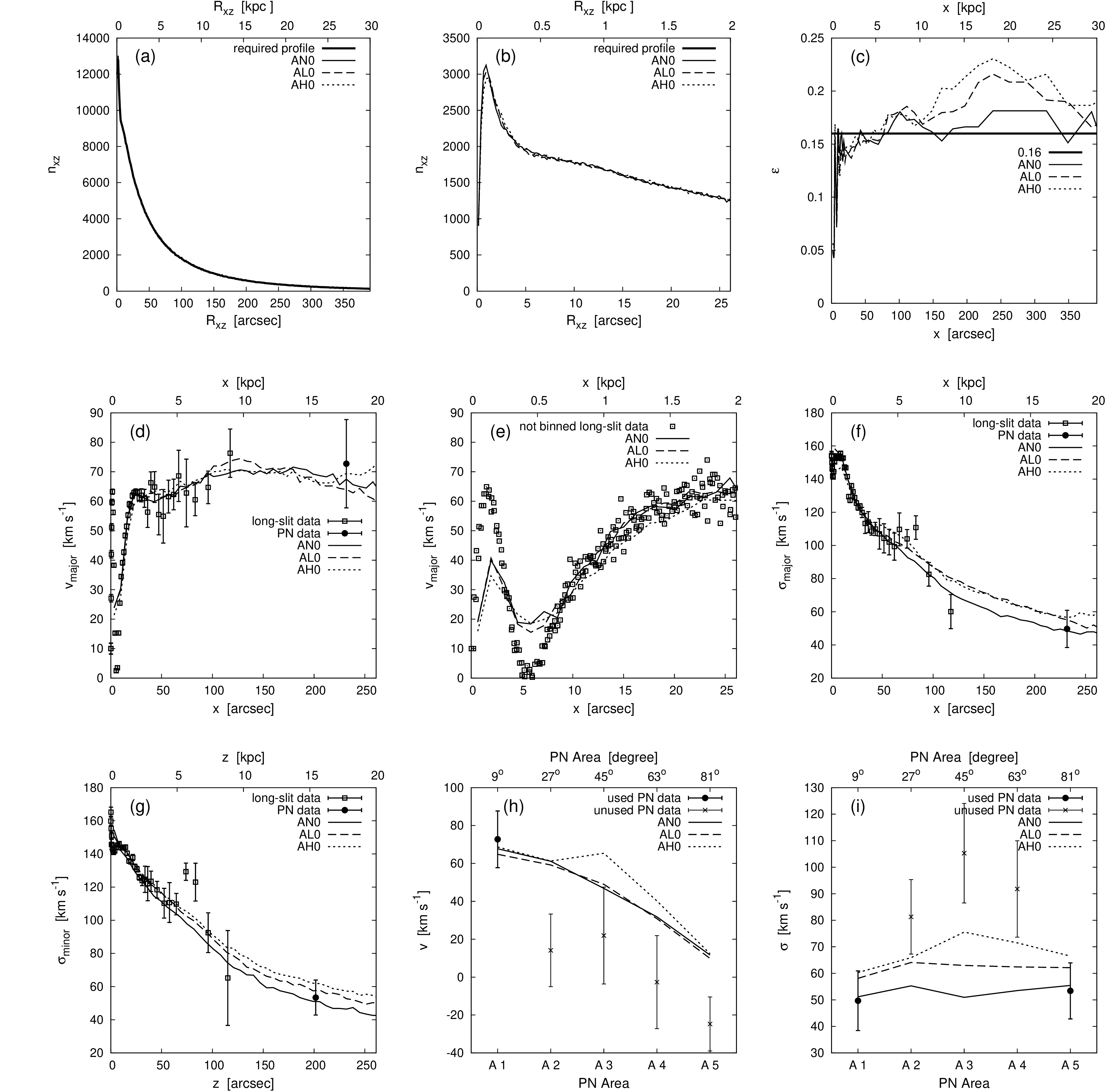}
\end{center}
\caption{Comparison of models AN0, AL0 and AH0 with the
observational data. (a) and (b) show the 
dependence of $n_{xz}$ on $R_{xz}$, where $n_{xz}$ is the number of
particles in concentric cylindrical shells and
$R_{xz}=\sqrt{x^2+z^2}$. The thick solid line shows the profile
calculated for a model with the input data prepared as described in
Sect.~\ref{s_photometry}. Panel (c) shows the ellipticity
profiles of the models projected on the sky plane (XZ plane), calculated with 
the IRAF task ELLIPSE. Panels (d) and
(e) show the profile of the mean velocity along the major axis and panels
(f) and (g) the  profiles
of the velocity dispersion along the major and minor axes, respectively. 
In (d) and (f), together with the
long-slit data, we show PNe data in area A1 which is close to the major axis 
(see Fig.~\ref{fig_pn_areas}). The value calculated for
this area is assigned to radius $(R_{xz}^{(min)} + R_{xz}^{(max)})/2$ 
(Table~\ref{t_pn}). In panel (g), together with the long-slit data, we
show the PNe data in area A5 which is close to the minor axis. 
In panels (d), (f) and (g) the observed data are binned as described in
Sect.~\ref{s_longslit}. Panel
(h) shows the mean line-of-sight velocity 
calculated for the five PNe
areas (see Sect.~\ref{s_pn}). Note that only the mean velocity in
area A1 is used for constructing the models
we use. Panel (i) shows the line-of-sight velocity
dispersion calculated for the five PNe areas. Only dispersions in areas
A1 and A5 are used for construction of A models. The radial profiles
for models AN0, AL0 and AH0 are given by solid, dashed and dotted
lines, respectively. The open squares show the long-slit data, not
binned for panel (e), and binned as discussed in Sect. ?? for the
remaining panels. The filled circles show the PNe data, which,
following N09, were used as observational constraints for these models,
and the X signs the remaining PNe data. } 
\label{fig_A0.p1}
\end{figure*}

\begin{figure}
\begin{center}
\includegraphics[width=8cm,angle=-90]{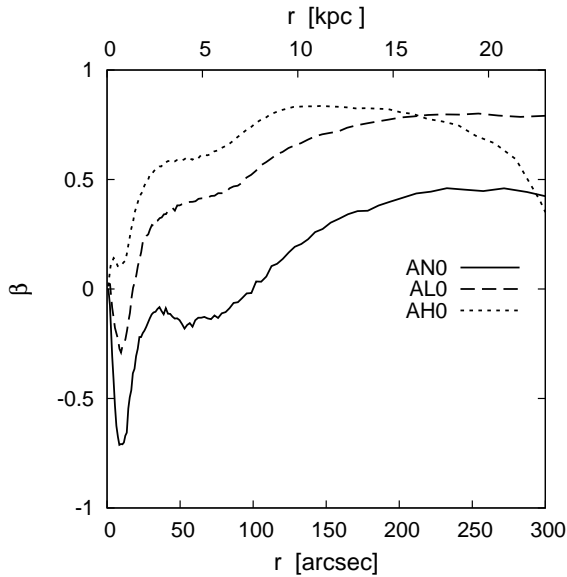}
\end{center}
\caption{
Radial profiles of velocity anisotropy for models AN0, AL0 and AH0. The
velocity anisotropy parameter is calculated as  
$\beta = 1 - \frac{\sigma^2_{\varphi}}{\sigma^2_{r}}$, where
$\sigma_{r}$ and $\sigma_{\varphi}$ are the velocity 
dispersion in the radial and the $\varphi$ direction, respectively, in
a spherical coordinate system.} 
\label{fig_A0.betta}
\end{figure}

\begin{figure*}
\begin{center}
\includegraphics[width=17.3cm,angle=-90]{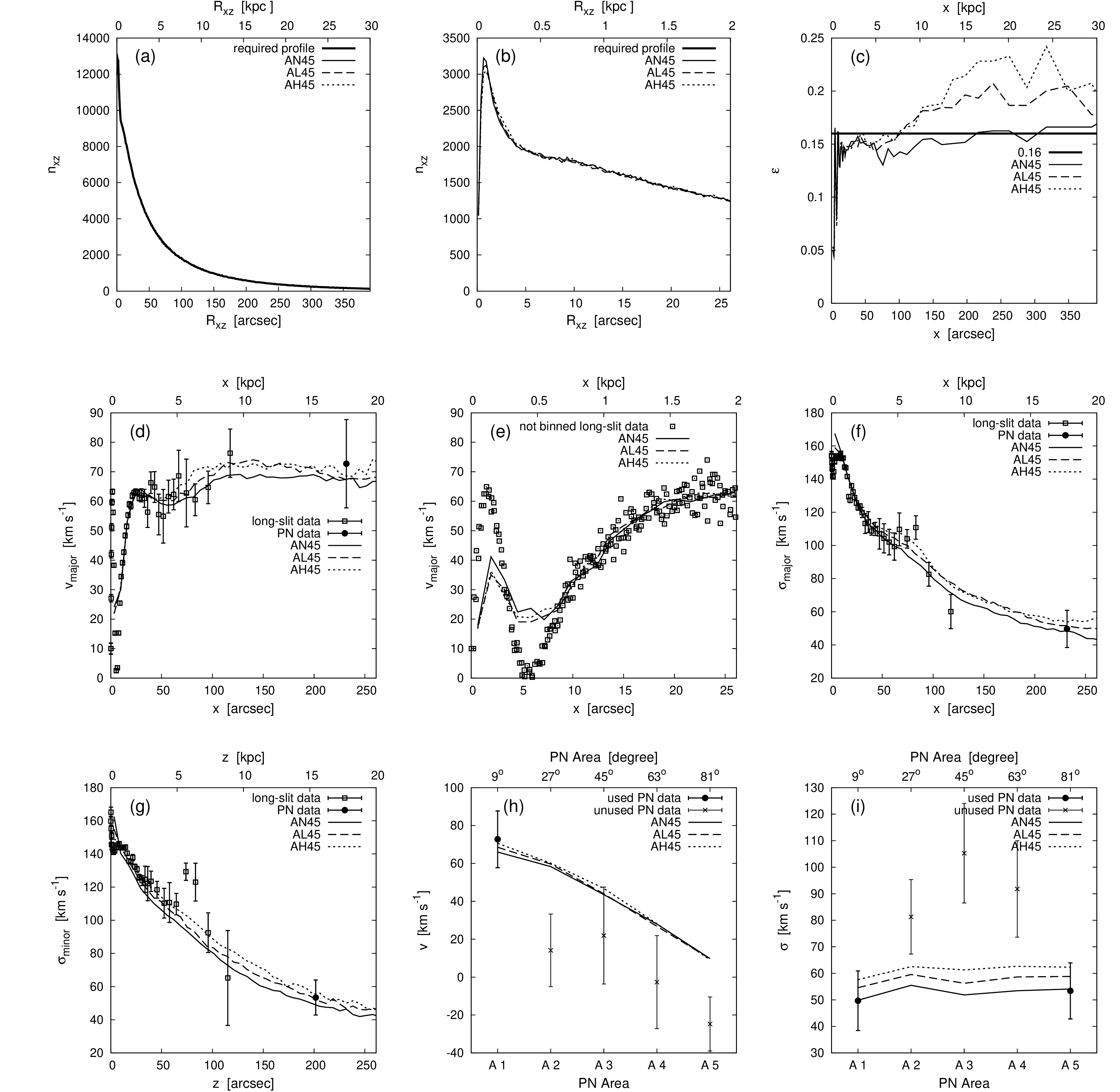}
\end{center}
\caption{As in Fig.~\ref{fig_A0.p1}, but for models AN45, AL45 and
  AH45.
}
\label{fig_A45.p1}
\end{figure*}

When constructing our first set of models, models A, we use
only velocity information from the vicinity of the 
principle axes. In particular, we take into
account the velocity dispersion in PNe areas A1 and A5 and ignore 
observations in areas A2, A3, A4. In so doing we want to demonstrate that our
method can be used to construct galactic models following
given observational constraints, both photometric and kinematic.

Let us first discuss models AN0, AL0 and AH0, which have
an inclination angle $\alpha=0$, i.e. their rotation axis is parallel to
the sky plane. Our results for these models are summarized in Table 2
and Figs.~\ref{fig_A0.p1} and \ref{fig_A0.betta}. In order to
increase the resolution and reduce the noise in our figures we use
a trick described by \citet{A05b}.  We stack ten snapshots closely spaced
in time, with $\Delta t = 10 \, \rm Myr$, and we calculate all values for this
combined snapshot. This allows us to reduce the noise in our plots
quite significantly. 


Since we include rotation, we compare models to observations
separately for mean velocities and for velocity dispersions, and do not fold
these two quantities into a single root-mean-square velocity 
$v_{\rm rms} = \sqrt{v^2 + \sigma^2}$ (N09).

In general, there is a good agreement between the models and the observational 
data which were used by the iterative method for constructing them
(Fig.~\ref{fig_A0.p1}). We get an excellent fit of the 
density profile (panels (a) and (b)). 
We also get good agreement with the mean velocities on 
the major axis except for radii within the centermost region 
(panels (d) and (e)) and
with the velocity dispersion on major and minor axes (except for the bump at 
radii between 60'' and 90'' -- see panels (f) and
(g)). All these agreements hold for all three models,
i.e. models AN0, AL0 and AH0. Also all our model fit very well mean velocity
of PNe in area A1 (panel (h)). 
Velocity dispersion in PNe areas will be discussed later in this section.
Let us now discuss in more detail the parts where the models fail to
reproduce the observations. 

\begin{itemize}
\item
As many other ellipticals, NGC 4494 has a kinematically decoupled
core. This is clear not only from the mean velocity, but also from the
velocity dispersion profiles along the major and minor axes. Such
features are believed to be out of equilibrium, e.g. due to a merger
with a a small 
companion that reached the galactic centre by dynamical
friction. Furthermore, the material in this region may not be  
axisymmetrically distributed and/or have a different orientation from
the remaining galaxy. Arguments in favour of these
possibilities are that the density distribution near the center
does not follow the S\'ersic law and that the profile of the
mean velocity does not have the shape expected for axisymmetric
mass distributions in equilibrium. 
Departures from axisymmetry and
from equilibrium are beyond the scope of this paper.
It is thus normal that neither N09 -- with a spherically symmetric
non-rotating equilibrium model -- nor we -- with an axisymmetric, 
rotating or non-rotating equilibrium model --
reproduce the structure in the innermost part. Models including
kinematically detached cores are beyond the scope of both studies.

\item
The velocity dispersion profiles have a bump along both the major and
minor axes in the area between $60''$ and $90''$ 
(panels (f) and (g) of Fig.~\ref{fig_A0.p1}).
N09 considered the long slit data as kinematical constraints only up
to $60''$, where they considered them as more accurate. But their plots show
clearly that their models have no such bump in the $60''$ to $90''$ region.
We kept all the long-slit
data as constraints, but none of our models reproduced this bump. 
It is possible that in NGC 4494 this
bump is transient, presumably due to some collision event
and thus can not be reproduced by the Jeans method, or by our iterative
method, which can only find equilibrium solutions. This, together with
the previous discussion on the centre-most area, argues that some regions of
NGC 4494 are not in exact steady state, as often observed and as could be
expected in the framework of the $\Lambda$CDM paradigm.

\item 
The ellipticity of model AN0 agrees on average with
the mean observed value (panel (c)). However models AL0 and AH0
have bigger ellipticity values than observed in their outermost
parts. This is partly due to the fact that these two 
models have slightly boxy isophotes at their periphery, which we
believe to be connected to the very high velocity anisotropy in
their periphery (Fig.~\ref{fig_A0.betta}),
i.e. to be linked to the radial orbit instability.  
As already mentioned, our algorithm for model construction includes
a self-consistent $N$-body evolution on a relatively long time scale 
(Fig.~\ref{fig_scheme}), so that our models can ``feel'' such
instabilities. Let us, however, underline that this effect
is relatively small and the isophote boxyness is rather subtle. As a
result the difference with the average ellipticity is of the order of
only 0.05, i.e. much smaller than the corresponding difference for
spherical models, such as those of N09. 
\end{itemize}

We can conclude that our  
models reproduce well the projected surface density, the mean apparent
ellipticity, the mean velocities and their dispersions (the latter
though not over the full radial and angular extent of the galaxy). 
The corresponding fits are no worse than the
N09 models. The discrepancies between our models and observations
occur mainly in regions whose data were not considered by N09. 

Fig.~\ref{fig_A0.betta} shows the radial profile of the velocity anisotropy
for our $\alpha$ = 0 models and clearly illustrates the degeneracy
between mass and anisotropy.
These three models have different halo masses, but the same stellar
density profile and observed
kinematics (Fig.~\ref{fig_A0.p1}) and this is possible because they
have different velocity anisotropies. As was expected, radial
anisotropy is increasing with halo mass (see Fig.~\ref{fig_A0.betta}). 

The inclined models AN45, AL45 and AH45 are considerably different
from the previously discussed AN0, AL0 and AH0 models, as they have an
intrinsic ellipticity of 0.35. The quality of the fits, however, is
very similar (Fig.~\ref{fig_A45.p1}). 

Seen the size of the error bars and the small effect of the halo mass
on the density and velocity radial profiles, it is not possible from
the above comparisons alone to rule out any of the models, or even to set a
strong preference to one rather than another. The only exception is
the ellipticity profiles, which are considerably better fitted by the
halo-less models AN0 and AN45 (Figs.~\ref{fig_A0.p1} and~\ref{fig_A45.p1}).
On the contrary, all models with dark halo have in their outer parts
slightly boxy isophotes (presumably due to their higher 
velocity anisotropy) and a higher than mean 
observed ellipticity. In our modeling, however, we use only the mean
value of the ellipticity and not the full profile. Also we don't use high order
isophote shape parameters, so we enforce our models to have precise
elliptical isophotes. NGC4494 has a negative $a_4$ isophote
shape parameter on periphery, i.e.  boxy isophotes (see appendix A in
N09). So we cannot rule out models with dark halos relying only on
the fact that they have slightly boxy isophotes in their outer parts.

Models without dark halo fit perfectly the velocity dispersion of the
PNe that have been used as observational constraints, i.e. in regions
close to the major and minor axes (see panels (i) of
Figs.~\ref{fig_A0.p1} and \ref{fig_A45.p1}).  
Model AL45, however, with a relatively light dark halo, also 
fits the observational data along the major and minor axes
rather well, while AL0 fits only slightly worst (see panels (i) of
Fig.~\ref{fig_A0.p1}~and~\ref{fig_A45.p1}).  
And, seen the error bars, even model AH45 with a relatively massive dark halo 
can not be ruled out. 

\subsubsection{Higher order moments of the velocity distribution} 
\label{subsub:highordermom}

\begin{figure*}
\begin{center}
\includegraphics[width=11.3cm,angle=-90]{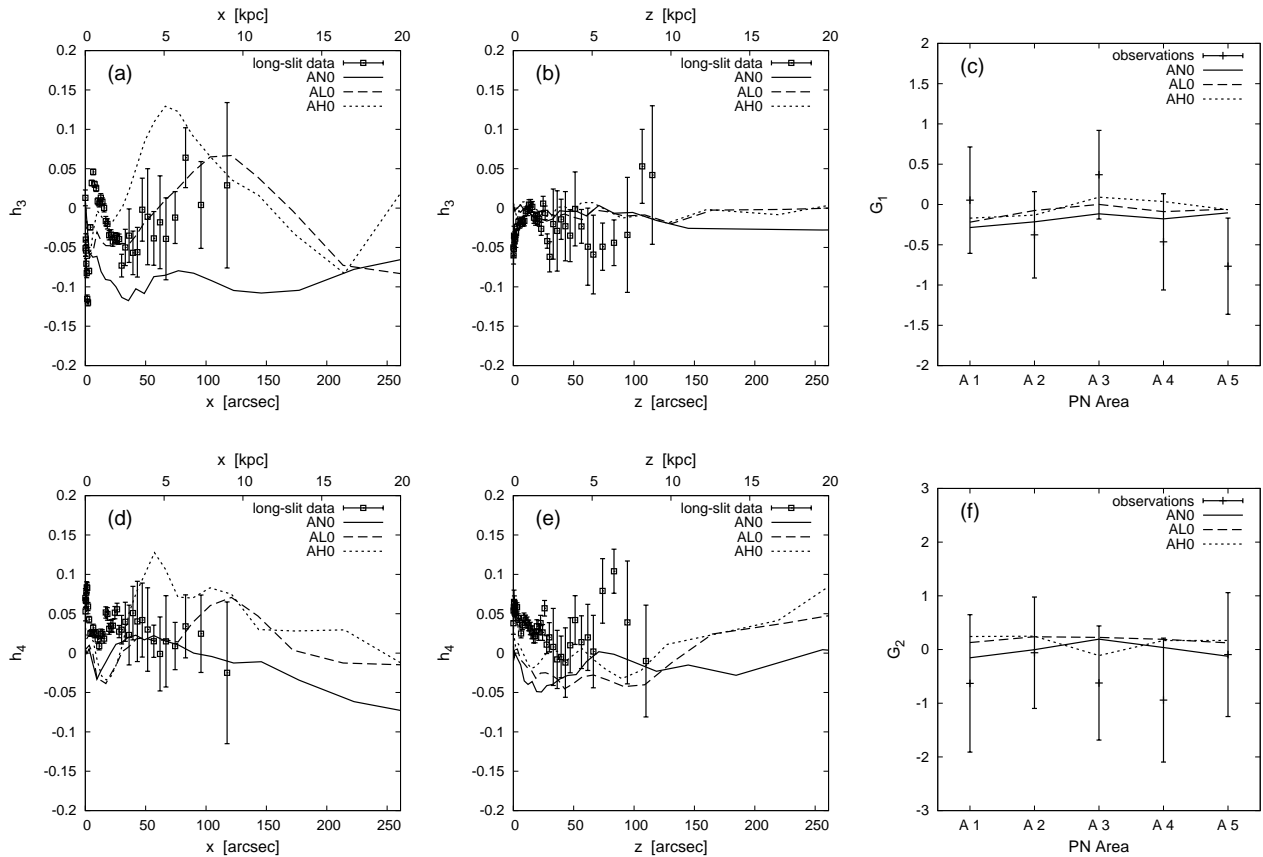}
\end{center}
\caption{
Comparison of higher moments of 
the velocity distribution of models AN0, AL0 and
AH0 with the observational data. Panels (a) and (b) show profiles of the
$h_3$ Gauss-Hermite moment along the major and minor axes,
respectively (calculated as described in appendix~\ref{s_gh}). Panel 
(c) shows an estimate of the skewness $G_1$ calculated for the five
PNe areas (calculated as described in appendix~\ref{s_sk}). (d) 
and (e) show profiles of the $h_4$ Gauss-Hermite moment along the
major and minor axes, respectively. (f) shows an estimate of the
kurtosis $G_2$ calculated for five PNe areas. These estimations of
the model skewness and the kurtosis are ``reduced'' to a
sample size equal to the number of PNe in the corresponding areas (see
appendix~\ref{s_sk}). The error bars in panels (c) and (f) were calculated
assuming a Gaussian distribution function (see
appendix~\ref{s_sk}). The symbols and line styles are as in
Fig.~\ref{fig_A0.p1}}.   
\label{fig_A0.p2}
\end{figure*}

\begin{figure*}
\begin{center}
\includegraphics[width=11.3cm,angle=-90]{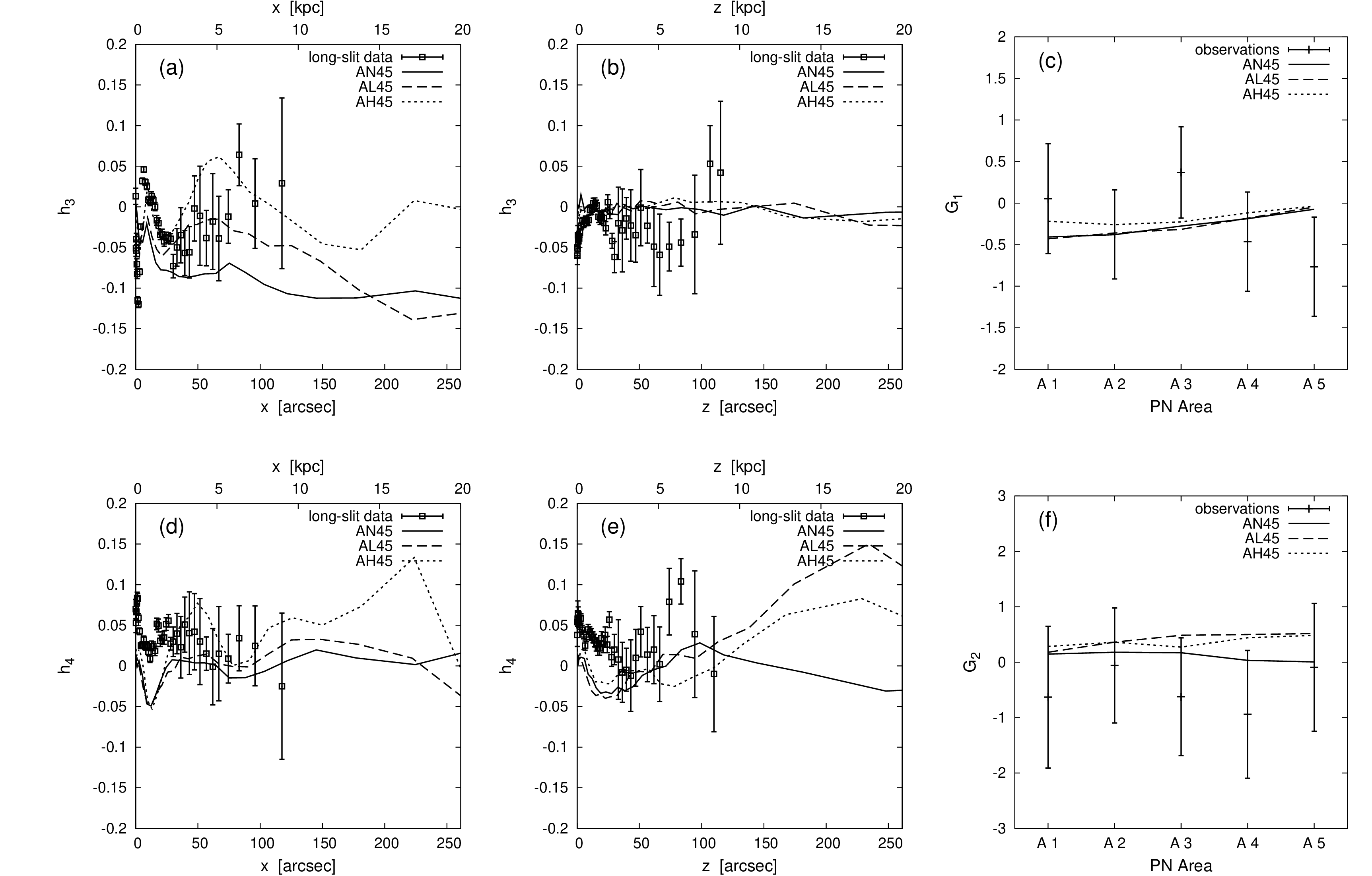}
\end{center}
\caption{Comparison of higher moments of 
the velocity distribution of models AN45,
AL45 and AH45 with observations. The values shown are the same
as in Fig.~\ref{fig_A0.p2}.}
\label{fig_A45.p2}
\end{figure*}

When we constructed the models, we did not use the higher-order
moments of the velocity 
distribution as observational constraints. It is thus interesting to compare
the third and fourth order moments of the velocity distribution of 
our models to those of the observational data. Moreover, such
moments may help break the mass-anisotropy degeneracy and were used by
N09 to argue for the need of a low mass halo. At the
periphery of the galaxy we have kinematical information
only from the PNe. 
We thus calculate the skewness and the kurtosis in the five PNe areas
(Fig.~\ref{fig_pn_areas}) and compare them to the corresponding
model values, calculated for our
models as described in appendix~\ref{s_sk}. It should, however, be noted that
the number of planetary nebulae is very small and that this affects
more the higher order moments. Thus, even 
assuming a Gaussian velocity distribution, the uncertainties
for skewness and kurtosis are very big, while without this assumption 
the uncertainties are formally infinite (appendix~\ref{s_sk}). 

One of the advantages of our models is that we can easily calculate
any parameters for them. This will allow us now to calculate for our
models exactly the same high-order moments of the velocity
distribution as for observations. We can thus calculate two
Gauss-Hermite moments ($h_3$ and $h_4$) along the major and minor axes
of the galaxy, as described in appendix~\ref{s_gh}, and compare them with
the corresponding profiles from long-slit spectroscopy. 
The results are given in
Figs.~\ref{fig_A0.p2} and \ref{fig_A45.p2} for $\alpha
= 0$ and $\alpha = 45$, respectively. These figures   
show clearly which of the higher order moments depend on the halo
mass and in which way. 

The most interesting of the Gauss-Hermit moment profiles is probably 
the $h_3$ profile along the major axis.
This is visibly different for models with different halo mass 
(see panels (a) of Figs.~\ref{fig_A0.p2} and~\ref{fig_A45.p2}), the most
massive halo one having the highest $h_3$ values and AN0 the
lowest. The best fit seems to be for the intermediate halo mass. To
establish this we calculated the $\chi^2$, excluding the region of the
kinematically decoupled core, and normalised it by the number of data
points. We find that for $\alpha = 0$, the values are 3.9, 0.4 and 5.7
for models with no halo, light halo and heavy halo, respectively. The
corresponding numbers for $\alpha = 45$ are 2.4, 0.8 and 1.8,
respectively. These numbers show a preference for the models with light
halo, and argue, albeit weakly, for a preference for $\alpha=0$.
The profile of $h_4$ along the
major axis also shows some dependence with halo mass, but much less so
than the
corresponding $h_3$ profile. (panels (d) of Figs.~\ref{fig_A0.p2} and
\ref{fig_A45.p2}). Profiles along the minor 
axis for both Gauss-Hermite moments show no clear dependence on
halo mass (panels (b) and (e) of
Figs.~\ref{fig_A0.p2} and \ref{fig_A45.p2}), except for  
the $h_4$ profiles in the outermost parts, where, however, there are
no long-slit data.
In the central part of the model, i.e. inside $\sim 30''$, all our models have
$h_4$ values which are clearly less than those of the observations for
both axes (Panels (a) of Figs.~\ref{fig_A0.p2} and~\ref{fig_A45.p2})
and this is true also for the $h_3$ major axis profiles. This could
again be linked to the kinematically decoupled core.

The kurtosis values calculated in each of the PNe areas for all ``A'' models 
are, for practical purposes, almost the same (panels (f) in
Figs. \ref{fig_A0.p2} and \ref{fig_A45.p2}), 
the differences being much smaller than the observational errors.
The situation with the skewness is the same (panels (c) in
Figs.~\ref{fig_A0.p2} and \ref{fig_A45.p2}).
We, therefore, do not believe that the
comparisons in panels (c) and (f) of Figs.~\ref{fig_A0.p2} and
\ref{fig_A45.p2} can be used to distinguish between models of
different mass.

To summarise, the $h_3$ profiles along the 
major axis provide some arguments in favour of models with a
light halo (particularly AL0 and AL45), in good agreement with what
was found by N09.


\subsection{Discussion of the second set of models}


\begin{figure*}
\begin{center}
\includegraphics[width=7cm,angle=-90]{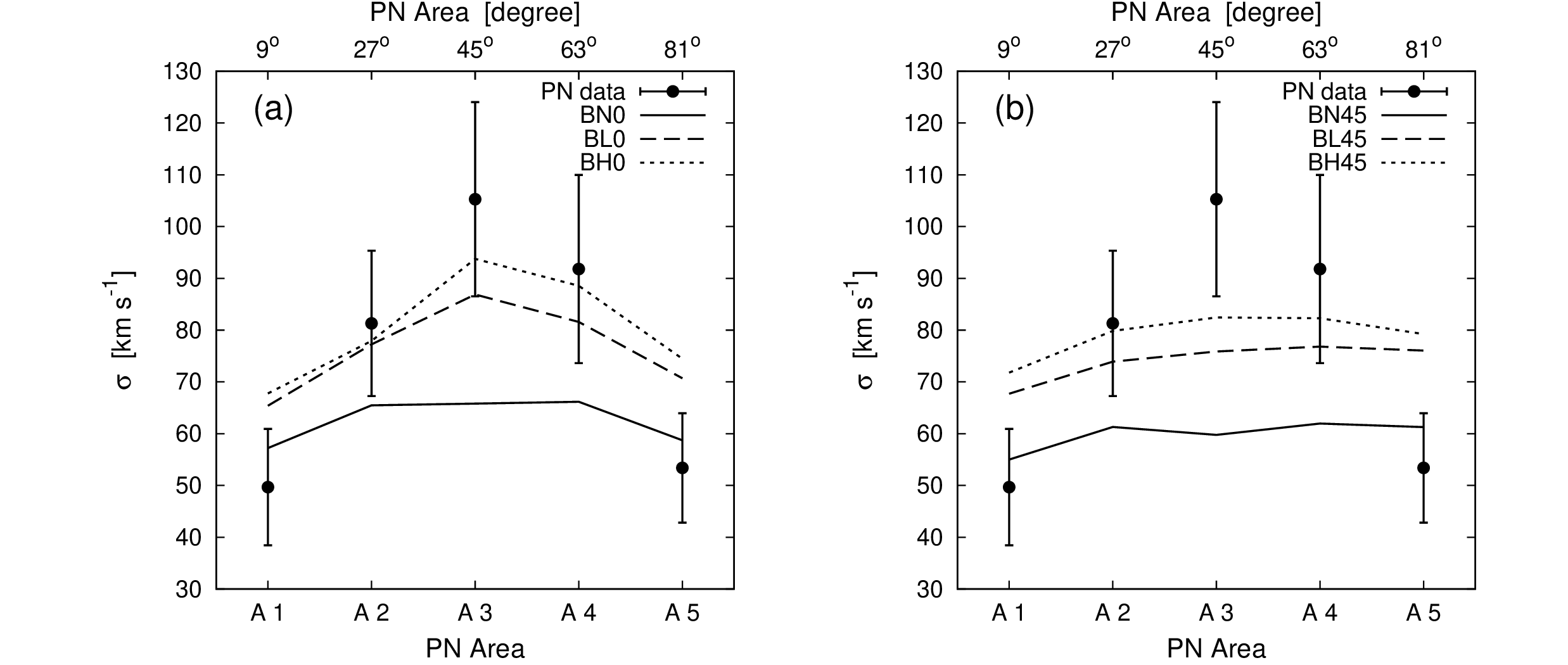}
\end{center}
\caption{Comparison of the line-of-sight velocity dispersion calculated for
the five PNe areas in the ``B'' models with observational data. Panel
(a) refers to models
BN0, BL0 and BH0 and panel (b) to models BN45, BL45 and BH45.} 
\label{fig_Bmodels}
\end{figure*}



As discussed in section~\ref{s_pn}, the velocity distribution of the PNe
has an interesting feature, namely that the velocity dispersion near
the principal axes (regions A1 and A5) has much smaller values than the
velocity 
dispersions in the intermediate areas (regions A2, A3 and A4). 
When building
the models presented in the previous section, we did not use this as an
observational constraint and ignored observations in intermediate areas.
As results all our ``A'' models 
have almost the same dispersion in all PNe areas (see panels (i) in
Figs.~\ref{fig_A0.p1} and \ref{fig_A45.p1}), so they obviously fail to
reproduce observational data in intermediate areas. Here we consider
more realistic models (``B'' models) for whose construction we used
all available observational data, to see whether the tangential variation of
the velocity dispersion can be reproduced by rotating axisymmetric
models.

In general both sets of models are very similar, with differences only 
at the periphery of the galaxy. 
Furthermore, the quality of the fit to the observation data is also similar.
Thus, to save space, we show only a very restricted set of figures. 
Fig.~\ref{fig_Bmodels} shows the velocity dispersion calculated in the
five PNe areas for our B models. 

We find that only non inclined models with a dark halo (i.e. models
BL0 and BH0) reproduce this feature, and even they only
partially. Namely, the velocity dispersions near the principal axes
(areas A1 and A5) are too high, while at the middle (region A3) it is
marginally too low. Thus we were able to reproduce qualitatively the
form of the azimuthal variation, but not quantitatively, since the
model amplitude is lower than necessary to fit the observations.
For these models, we can
not make specific comparisons with N09, because the latter did not try
and reproduce this feature. 

We can thus conclude that, although our models reproduce most of the
observed features well, they fail to fully reproduce the velocity 
distribution of planetary nebulae. 
We will discuss possible explanations in the next section.


\section{Conclusions}
\label{s_conc}

We used our iterative method \citep{RAS09} to construct dynamical models of
NGC 4494. One of the advantages of our method is that the models are
produced in the form of
$N$-body snapshots with equal mass particles, which can be directly used
in $N$-body simulations. This, for example, allows us to check directly
whether the constructed model is in steady state and to
calculate any quantities, or parameters of these models, which can
then be compared to observations. As already discussed in
Sect.~\ref{s_method} (see also Fig.~\ref{fig_scheme}),  
after the main part of the iterative method, we 
let our models evolve freely on a time scale of 3 Gyr, and then only
do we calculate the quantities to be compared with the observational
data. Since the aim of this evolution is to check
whether the models we constructed are indeed in steady state, we used
the same halo as when building the model, i.e. a rigid halo.   
This evolution showed clearly that all our models are
in a steady state. For completeness sake, however, we also tried
simulations with a live halo and we find no considerable differences
from the rigid halo simulations, arguing that in these cases there is
not much interaction between the dark and the stellar matter. 

We used the observational data given by N09, namely surface
photometry, stellar kinematics along the major and minor axes as obtained by
means of long-slit spectroscopy, and velocities and positions of 255 planetary
nebulae (see N09). We use PNe data only in the outer part of
the galaxy where long-slit data are absent. The surface photometry
gives us the surface distribution of 
particles but not the total mass of the system because the mass-to-light
ratio is unknown. Our algorithm automatically adjusts the total mass of the
model (Sect.~\ref{s_method}). 

We constructed models with three types of halo, all of which were
already discussed in N09. The first type is models without halo.
The second type of halo is a relatively light NFW halo, which N09 found 
to be the best-fitting NFW for NGC 4494. The third type of halo is a relatively
heavy NFW halo. We, furthermore, constructed two
sets of models: models ``A'' and models ``B''. These two sets have only one
difference. When constructing models ``A'' we
used as observational constraints only the PNe in areas close to the
principal axes, neglecting  
information on the velocity distribution of the PNe in intermediate areas.
on the contrary, in order to construct models ``B'' we used all
available information. 

One important goal in making the ``A'' models was 
to demonstrate the ability of our method to construct equilibrium
models with given observation constraints. This was fully achieved.
In general, there is a good agreement between our
models and the observational data concerning the projected surface
density, the mean apparent ellipticity, the mean velocities and their
dispersions (except for specific radial ranges, where some
non-equilibrium or non-axisymmetric structure  
could be present in NGC 4494). We showed that our models reproduce
observations data not worse than N09 models. But our models have the
added advantage that they are live, non-spherical,
rotating $N$-body systems. 

A further goal was to see whether it was
possible to set constraints on the dark halo mass with our models. 
If we don't take into account the high-order
moments of the velocity distribution, then the best models are the models
without dark halo. This, however, is only a slight preference and models with
a light halo and even models with a heavy halo can not be rejected. It
is thus necessary to try higher order moments. 
We found that the major axis
profile of the third order moments shows the strongest dependence on the halo
mass and that the best fit was for the
models with the light halo. This is in agreement with what N09 found for their
models, though with a different technique.

N09 found that a heavy halo gives worst fits to the data, without referring
to higher order moments, i.e. from the root-mean-square velocity
profile, which is a combination of the mean velocity and the
dispersion. This, however, is not the case for our models. For example,
model AH45, which has a heavy halo, fits all low order moments of 
the velocity distribution very well. These two results taken together,
show that there are no simple spherical models fitting the lower
order models, but that there can well be more general models that
do. Thus, our model which is axisymmetric, rather than spherical, rotating
and inclined with respect to the line of sight, has no difficulty with
the lower order moments. When we place the bar higher,
i.e. when we ask our models to fit also the third and fourth order
moments, we find that the models with heavy halos do worse
than the models with lower mass halos, as we described in Sect. 
\ref{subsub:highordermom}. This does not necessarily mean that there
are no heavy halo 
models that fit the third and fourth order moments. It can simply mean
that it is necessary to consider a yet more general model, e.g. a
triaxial one, in order to get such a fit. In other words, one has to
be careful not to extrapolate a given result further than the class of
models for which it was derived. And this of course makes it very
difficult to set any strong constraints to the halo mass in NGC 4494,
and to ellipticals in general.

N09 found that some halo is required by comparing the kurtosis of
their model with that of the PNe velocity distribution at the
outermost parts of NGC 4494. Again, this is not the case for our
models. It is the $h_3$ radial profile along the major axis that
proved in our case to be more sensitive to the halo mass, and thus
allowed us to set some preference for the light halo model, in
agreement with what was found by N09 for the fourth order moment.
 
When constructing and analysing the ``A'' models we ignore the PNe
data at angles intermediate between the major and the minor axes.
Since this is rather ad hoc,
we built the ``B'' models, in which all PNe data were used as
observational constraints. By doing so, we wanted to test whether we
could make models which reproduced the azimuthal variation of the
velocity dispersion, i.e. which have relatively low values near the
principal axes and considerably higher ones in between. Although we
were able to built models that qualitatively reproduce this feature,
we were unable to reproduce it quantitatively; i.e our models always
display a lower amplitude of this variation than the observations.  
There are several possible explanation to this.

It should first be noted that the observational data at the periphery
of NGC 4494 is very sparse. We have only around 15 PNe in each of our
PNe areas (see Table~\ref{t_pn}). So even the formal uncertainties of
the PNe dispersions in these areas are rather big. Moreover, there
could still be some contamination in the PNe sample (see section 2.2
in N09), which would further increase the uncertainties. As a result, there 
is still a considerable probability that the observations do not
contradict e.g. model BL0 (see fig.~\ref{fig_Bmodels}.a,
error-bars are one $\sigma$ ). 

There is of course always the possibility that a model which fully
reproduces the observations exists, but that our method failed to construct it.
We made
very extensive searches and we do not believe that this possibility is
likely, but, strictly speaking, we can not exclude it. Furthermore,
the observations point 
to other, more likely alternatives, which we discuss below.

\begin{itemize}
\item
In our models the halo is spherical. It is likely that 
models with an axisymmetric or a triaxial halo would reproduce
observations better. Note, however, that this alternative cannot help
the models without a dark halo. 
\item
Our models are axisymmetric, while, as discussed in Sect.~\ref{s_pn},
the velocity distribution in the galaxy is not fully 
axisymmetric. 
It would thus have been better to 
consider more general types of models, for example, triaxial models or models
with even less symmetry. 
This, however, would introduce further free parameters and
is beyond the scope of this paper.
\item
It is possible that the outer parts of NGC 4494 are not in a steady
state. 
As we already discussed, there is evidence that NGC 4494 is not 
exactly in steady state. And if there are transient features in 
the innermost and perhaps in part some of intermediate part of the
galaxy (the kinematically decoupled core and between $60''$ and $90''$), then 
such features can also be present at its periphery. Moreover, the
dynamical time scale at the periphery is much longer than 
in the intermediate part, so that any non-equilibrium structures  
will tend to equilibrium much slower than
in regions further in. 
\item
Seen the low number of PNe in the periphery of the galaxy, the
uncertainies are very large and therefore the constraints in these
regions very loose. It might therefore not be necessary to dismiss a
model if the fits in these regions are poor.
\end{itemize}

Any of the three first alternatives would necessitate a more general
model 
than what we have considered here. Triaxial models
could, at least in principle, be built, but the number of free
parameters would increase so that more observational constraints would
be necessary. However, neither our techniques, nor other
techniques discussed so far, can build non-equilibrium models. This
could be done only by $N$-body simulations, which,
however, would have an extraordinarily unwieldy free parameter 
space.

To summarise, we were able to reach our first goal, namely we showed 
that our iterative method (RAS09) can be used for building models
with given observational constraints. Concerning our second goal,
i.e. to set strict constraints on the halo mass of NGC 4494, we can
only claim a more limited success. Comparing the third order moments of
our model velocity distribution with the observations, we find that
it is the light halo model that gives the best fit. Even this,
however, does not fit all available data, and it could be that a more
general model (e.g. triaxial) is necessary. It is, however, not
possible to extend the preference of light halos to these more general
models, without actually making them and analysing their
properties. It could thus well be that it is another type of model
(such as with no halo, or with a heavy halo) that gives the best fits
in such a case. Thus the present set of models, although showing a
preference for a light halo, can not set a very strict
constraint to the halo mass of NGC 4494, particularly seen the
sparseness of the PNe in the outermost parts.

\section*{Acknowledgments}
We thank Albert Bosma for useful discussions on the data. 
This work was partially supported by grant ANR-06-BLAN-0172,
by the Russian Foundation for
Basic Research (grants 08-02-00361-a and 09-02-00968-a)
and by a grant from the President of the Russian
Federation for support of Leading Scientific Schools (grant NSh-1318.2008.2).

\bibliography{article_2009_2}

\appendix

\section{Calculation of Gauss-Hermite moments in the case of $N$-body systems}
\label{s_gh}

Let us now describe how we calculate the Gauss-Hermite moments
in the case of $N$-body systems. 
Both \citet{vdM93} and \citet{G93}, independently, discuss the use
of Gauss-Hermite moments to measure the deviations of the
observed velocity distribution from a Gaussian, but with different
normalization of the Hermite polynomials. Here we use the
normalization of \citet{G93}.

 The Hermite polynomials are defined as
\begin{equation}
H_n(x)=(-1)^n\, e^{ x^2} \, \frac{d^n}{dx^n}\left( e^{ -x^2}\right)
\end{equation}
The sequence of Hermite polynomials satisfies the recursion
\begin{equation}
H_{n+1} = 2 x H_n - 2 n H_{n-1}
\end{equation}
and the first three Hermite polynomials are
\begin{equation}
\label{eq_Hfirst}
H_0 = 1, \; \; \; H_1 = 2x, \; \; \; H_2 = 4 x^2 - 2. 
\end{equation}
The set of functions defined as
\begin{equation}
\label{eq_u}
u_n(x) = (2^{n+1} n! \pi)^{-1/2} H_n(x) \exp(-x^2/2) \, ,
\end{equation}
obey the orthogonality relation
\begin{equation}
\int^{+\infty}_{-\infty} u_n(x) u_m(x) dx = \frac{\delta^n_m}{2 \pi^{1/2}}.
\end{equation}
Thus this set of functions is a complete orthogonal system. 

The Gauss-Hermite moments for some function $l(v)$ are defined as
\begin{equation}
h_n = 2 \pi^{1/2}\gamma_h^{-1} \int^{+\infty}_{-\infty} l(v) u_n(w) dv \, , 
\; \; \; w \equiv (v - V_h)/\sigma_h     
\end{equation}
where  $V_h$, $\sigma_h \neq 0$ and $\gamma_h \neq 0$ are free parameters.
The function $l(v)$ can be approximately calculated by means of a truncated
Gauss-Hermite series
\begin{equation}
l(v) \approx \frac{\gamma_h}{\sigma_h} \sum^{N_h}_{i=0} h_i u_i(w), 
\end{equation}
where $N_h$ is the number of terms used (see \citet{G93} and \citet{vdM93} for
more details).

 In the case of an $N$-body system we need to solve the following problem. We
have the set of values $v_1$, $v_2$ {\ldots}  $v_{N}$, and we need to
calculate the Gauss-Hermite moments of the distribution function $l(v)$ 
of these values.
For example, if we need to calculate Gauss-Hermite moments for the
line-of-sight velocity distribution in some area of an $N$-body system 
then $v_i$ is the line-of-sight velocity of each particle in this area.
In this case, the Gauss-Hermite moments for the
function $l(v)$ can approximately be
calculated as
\begin{equation}
\label{eq_ghd}
h_n = \frac{2 \pi^{1/2}}{\gamma_h N} \sum^{N}_{i=1} u_n(w_i) \, , 
\; \; \; w_i \equiv (v_i - V_h)/\sigma_h .  
\end{equation}
We use this equation as the definition of Gauss-Hermite moments 
for our discrete
case.

To compare Gauss-Hermite moments with the observations, we need to
choose the free
parameters $V_h$, $\sigma_h$ and $\gamma_h$ as observer would do. These free
parameters should be chosen so as to give $h_0 = 1$, $h_1=h_2=0$ \citep{vdM93}.
From (\ref{eq_Hfirst}), (\ref{eq_u}), (\ref{eq_ghd}) for $n=1$ and the
condition $h_1 = 0$, we have
\begin{equation}
\label{eq_Vh}
h_1 = 0 \; \Leftrightarrow \; V_h = \displaystyle\frac{\sum^{N}_{i=1} v_i \exp(-w_i^2 / 2)}
{\sum^{N}_{i=1} \exp(-w_i^2 / 2)}    
\end{equation}
From (\ref{eq_Hfirst}), (\ref{eq_u}) (\ref{eq_ghd}) for $n=2$ and the
condition $h_2 = 0$, we have 
\begin{equation}
\label{eq_sh}
h_2 = 0 \; \Leftrightarrow \; \sigma_h^2 = 
2 \displaystyle\frac{\sum^{N}_{i=1} (v_i - V_h)^2 \exp(-w_i^2 / 2)}
{\sum^{N}_{i=1} \exp(-w_i^2 / 2)}.    
\end{equation}
We note that $w_i \equiv (v_i - V_h)/\sigma_h$, so (\ref{eq_Vh}) and
(\ref{eq_sh}) cannot be solved directly. We solve (\ref{eq_Vh}) and
(\ref{eq_sh}) together, by means of iterations. Initially we set $V_h$ equal
to the mean value of $v_i$, $\sigma_h$ equal to 
the standard deviation of $v_i$ and
$\gamma_h=1$. A single iteration is as follows

\begin{equation}
\begin{array}{rcl}
V_h^{(new)} &=& \displaystyle\frac{\sum^{N}_{i=1} v_i \exp(-w_i^2 / 2)}
{\sum^{N}_{i=1} \exp(-w_i^2 / 2)} \\
\sigma_h^{(new)} &=& 
\left(2 \displaystyle\frac{\sum^{N}_{i=1} (v_i - V_h)^2 \exp(-w_i^2 / 2)}
{\sum^{N}_{i=1} \exp(-w_i^2 / 2)} \right)^{1/2}.
\end{array}
\end{equation}
After finding the appropriate $V_h$ and $\sigma_h$ values, we can easily 
calculate the last free parameter as $\gamma^{new}_h = h_0$.

\section{Calculation of skewness and kurtosis for a small sample}
\label{s_sk}

In the outer parts of the galaxy we have information about the line-of-sight 
velocity distribution only from PNe observations. Our task
is to compare the velocity distribution of the PNe in some area on the sky
plane with the velocity distribution in a constructed $N$-body model (see
Fig.~\ref{fig_pn_areas}). More precisely our task can be formulated
as follows. 
We have two random samples which we denote here as ``A'' and ``B''. 
Sample~``A'' with size $n_a$ consist of the line-of-sight velocities
of the PNe in the selected area. Sample~``B'' with size $n_b$ consist of 
the line-of-sight velocities of model particles in the same area.  
We need to assess the probability that these
two samples were generated from the same distribution function. 
This can be achieved by comparing moments calculated for these two samples. 
 
Here we will discuss the comparison of the 
skewness and kurtosis calculated for
these two samples.
In our case the number of planetary nebulae in a given
area (the size of the sample~A) is small, not exceeding 18 
(Table~\ref{t_pn}). On the other hand the size of sample~B, i.e. the
number of particles in given area of constructed model is rather large, 
and is about $10^4$. So we need to compare high-order moments calculated for 
small and for large samples.  
Of course, for a sample as small as A the uncertainties of the estimators of 
the higher-order moments are rather large. Moreover, we cannot calculate these
uncertainties without making an assumption about the distribution function. 
If we assume, for example, a Gaussian
distribution then we can calculate these uncertainties \citep{J98}, but if
we do not assume a priori any distribution function then the
uncertainties are formally infinite. Furthermore, 
in the general case all commonly used estimators of the sample
skewness and kurtosis are biased \citep{J98}. For small samples
this bias can be very large, especially for the kurtosis (see tables 2 and 3
in \citet{J98}). 
Consequently, for sample~A the bias can be rather large.
The bias depends on the size of the sample and for sample~B (which is 
rather large) 
the bias is negligibly small. If both of our samples are
generated from the same distribution function then the expected values of
kurtosis (and skewness) for these two samples can differ significantly.
We can solve this problem reducing sample~``B'' to a smaller sample size. 

Let us now describe how we calculate kurtosis. The skewness is calculated in
the same way. Let us have some estimator of sample kurtosis. For sample~A we
simply 
calculate value of kurtosis $K_a$ using the chosen estimator. For sample~B
we calculate the value of the kurtosis ``reducing'' the sample size to
that of sample A, i.e. $n_a$. We randomly get 
$n_a$ members from sample~B and calculate for this sub-sample the kurtosis
$k_1$. We repeat this $N_k$ times and calculate
$k_1$, $k_2$ {\ldots}  $k_{N_k}$ 
(in this article $N_k = 100\,000$ ). We denote the mean value of $k_i$ as
$K_b^{n_a}$ and the standard deviation of $k_i$ as
$\sigma_b^{n_a}$. Let $f_b(x)$ 
be the distribution function corresponding to sample~B. For the chosen
kurtosis estimator we can construct the
distribution function $g_b(x)$ of the sample kurtosis for 
samples with size $n_a$ generated 
from $f_b(x)$. We note that $K_b^{n_a}$ is approximately equal to 
the expected
value of $g_b(x)$ and $\sigma_b^{a}$ is approximately equal to the standard
deviation of $g_b(x)$. For example, if the sample~A was also generated from the
distribution function $f_b(x)$, then $K_b^{n_a}$ would be the expected value 
and $\sigma_b^{n_a}$ the standard deviation of the kurtosis of sample~A. 

We use $K_a$ as measure of the kurtosis of sample~A, $K_b^{a}$ as measure
of the kurtosis of sample~B, and $\sigma_b^{a}$ as measure of the standard
deviation. We note that the main value for us is 
$\tau=|K_a-K_b^{a}|/\sigma_b^{n_a}$. \citet{J98} discussed three different
estimators of the kurtosis. 
For our analysis, however, it doesn't matter which of the three we use
because 
it can be proven that the value of $\tau$ is the same for all of them.
To calculate the sample skewness and the
sample kurtosis we use the estimators $G_1$ and $G_2$, discussed in \citet{J98}.
These estimators are unbiased in the case of a normal distribution. 

As we noted above, we cannot calculate uncertainties of estimators of
kurtosis or skewness without making an assumption about the distribution
function. The value $\sigma_b^{a}$ is the standard deviation of the estimation
of the kurtosis or the skewness for sample A (observation) assuming
that it s from the same 
distribution function as sample B (model). So this value depends on the model. 
We also can calculate the standard deviations $\sigma_{n}^{a}$
of the estimation of kurtosis or skewness for sample A 
assuming that its distribution function is Gaussian \citep{J98}. For our
models in general value 
of $\sigma_{n}^{a}$ is less that of $\sigma_b^{a}$. The difference, however, 
is not so strong. The maximum difference is 
$\sigma_b^{a}/\sigma_{n}^{a} \approx 1.3$ and 
$\sigma_b^{a}/\sigma_{n}^{a} \approx 1.4$ for the skewness and the
kurtosis, respectively.
\label{lastpage}
\end{document}